\documentclass[twocolumn]{aa} 

\usepackage[nolist]{acronym}

\usepackage{graphicx}
\usepackage{threeparttablex}
\usepackage{txfonts}

\usepackage{amsmath,amsfonts,amssymb}
\usepackage{bm}   
\usepackage{graphicx}
\usepackage[colorlinks=true, allcolors=blue]{hyperref}
\usepackage{astro_bib_macro}
\usepackage{ulem}

\def\Var{{\textrm{Var}}\,}

\DeclareMathOperator{\tr}{tr}

\usepackage{booktabs}

\usepackage{listings}

\graphicspath{{./}{./figures/}}   

\makeatletter
\renewcommand*\aa@pageof{, page \thepage{} of \pageref*{LastPage}}
\makeatother

\begin{document}

\title{Wavefront tolerances of space-based segmented telescopes at very high contrast: Experimental validation}

\titlerunning{Wavefront stability of segmented telescopes}
\authorrunning{Laginja \emph{et al.}}

\author{Iva Laginja\inst{\ref{inst-a}}\inst{\ref{inst-b}} \and Jean-Fran\c{c}ois Sauvage\inst{\ref{inst-a}}\inst{\ref{inst-b}} \and Laurent M. Mugnier\inst{\ref{inst-a}} \and Laurent Pueyo\inst{\ref{inst-c}} \and Marshall D. Perrin\inst{\ref{inst-c}} \and James Noss\inst{\ref{inst-c}} \and Scott D. Will\inst{\ref{inst-c}}\inst{\ref{inst-d}} \and Keira J. Brooks\inst{\ref{inst-c}} \and Emiel H. Por\inst{\ref{inst-c}} \and Peter Petrone\inst{\ref{inst-c}}\inst{\ref{inst-e}} \and R\'{e}mi Soummer\inst{\ref{inst-c}}}

\institute{DOTA, ONERA, Universit\'e Paris Saclay, F-92322 Ch\^{a}tillon, France\label{inst-a}
\and
Aix Marseille Universit\'{e}, CNRS, CNES, LAM (Laboratoire d'Astrophysique de Marseille) UMR 7326, 13388 Marseille, France\label{inst-b}
\and
Space Telescope Science Institute, Baltimore, MD 21218, USA\label{inst-c}
\and
The Institute of Optics, University of Rochester, Rochester, NY 14627, USA\label{inst-d}
\and
Hexagon Federal, Chantilly, VA 20151, USA\label{inst-e}\\
\email{iva.laginja@lam.fr}
}

\date{Received September 3, 2021; accepted October 21, 2021}

  \abstract
  {The detection and characterization of Earth-like exoplanets (exoEarths) from space requires exquisite wavefront stability at contrast levels of $10^{-10}$. On segmented telescopes in particular, aberrations induced by cophasing errors lead to a light leakage through the coronagraph, deteriorating the imaging performance. These need to be limited in order to facilitate the direct imaging of exoEarths.}%
  {
    We perform a laboratory validation of an analytical tolerancing model that allows us to determine wavefront error requirements in the $10^{-6} - 10^{-8}$ contrast regime, for a segmented pupil with a classical Lyot coronagraph. We intend to compare the results to simulations, and we aim to establish an error budget for the segmented mirror on the High-contrast imager for Complex Aperture Telescopes (HiCAT) testbed.
  }
   {
   We use the Pair-based Analytical model for Segmented Telescope Imaging from Space (PASTIS) to measure a contrast influence matrix of a real high contrast instrument, and use an analytical model inversion to calculate per-segment wavefront error tolerances. We validate these tolerances on the HiCAT testbed by measuring the contrast response of segmented mirror states that follow these requirements.
   }
   {
   The experimentally measured optical influence matrix is successfully measured on the HiCAT testbed, and we derive individual segment tolerances from it that correctly yield the targeted contrast levels. Further, the analytical expressions that predict a contrast mean and variance from a given segment covariance matrix are confirmed experimentally.
   }
   {}
   

   \maketitle
%

\section{Introduction}
\label{sec:introduction_exp_paper}

The search for Earth-like exoplanets (exoEarths) and potential signs of life in the form of atmospheric biomarkers is a very exciting field of today's astronomy. However, it requires a tremendous improvement in imaging capabilities compared to what we can currently achieve, in order to capture the few photons coming from a planet buried in the blinding light from its nearby host star. Instruments will need to reach contrast levels (planet to star flux ratios) of at least $10^{-10}$, at a separation of only $\sim 0.1$ arcsec, or less, from the star \citep{TheLUVOIRTeam2019LUVOIRFINALREPORT}. Telescopes providing these capabilities will require large collecting areas, and they will most likely be realized with segmented primary mirrors, both in space and on the ground. Currently, the favored method to achieve these extreme high contrast levels are dedicated instruments called coronagraphs that strongly attenuate the on-axis star light while preserving the off-axis planet light as much as possible \citep{Guyon2006TheoreticalLimitsExtrasolar}. These instruments are very sensitive to residual wavefront aberrations, which generate speckles of light in the imaging focal plane that can be mistaken for planets. This is why coronagraphy needs to be combined with wavefront sensing and active control (WFS\&C) \citep{Mazoyer2018a:ActiveCorrectionApertureI,Mazoyer2018b:ActiveCorrectionApertureII,Groff2016MethodsLimitationsFocal} to create a zone of deep contrast in the final image, a dark hole (DH). These ambitious goals will be achieved from space by missions such as the Habitable Exoplanet Observatory \citep[HabEx;][]{Gaudi2019HabitableExoplanetObservatory} and Large UV Optical InfraRed Surveyor \citep[LUVOIR;][]{TheLUVOIRTeam2019LUVOIRFINALREPORT, Bolcar2019LargeUVOptical} currently under consideration by the NASA Astro2020 Decadal Survey, with the Nancy Grace Roman Space Telescope \citep[RST;][]{Krist2015OverviewWFIRSTAFTA} working towards shorter-term demonstrations at more moderate contrast levels ($10^{-7} - 10^{-9}$) with a monolithic primary mirror.

The extreme contrast levels that are needed for the detection of exoEarths require excellent stability against wavefront errors (WFE) over a range of temporal and spatial frequencies \citep{Pueyo2019LUVOIRExtremeCoronagraph,Coyle2019LargeUltrastableTelescope,Feinberg2017UltrastableSegmentedTelescope}. While some of these can be actively controlled with a WFS\&C system, or do not have a large impact on the contrast due to robust coronagraph designs, aberration modes to which the instrument is very sensitive must be held to a minimal level \citep{Nemati2020MethodDerivingOptical,Juanola-Parramon2019LUVOIRExtremeCoronagraph,Nemati2017EffectsSpaceTelescope}. Typically, it is enough to control these misalignment modes on the time scales of the WFS\&C system. In this paper, we focus on the segment-related aberrations due to segment misalignments, which are the main contributors to WFE in the mid-spatial frequency regime \citep{Douglas2019LaserGuideStar,Juanola-Parramon2019LUVOIRExtremeCoronagraph,Moore2018PicometerDifferentialWavefront}. Various technology solutions are being developed towards this goal \citep{Coyle2020ProgressHardwareDemonstrations,Hallibert2019TechnologiesLargeUltrastable,Coyle2018EdgeSensorConcept,East2018PicometerLevelStability,Stahl2017AdvancedMirrorTechnology,Stahl2015PreliminaryAnalysisEffect}. Multiple hardware efforts are underway to provide laboratory demonstrations of the systems anticipated to be installed on future large observatories. While other testbeds are tackling the problem on monolithic apertures \citep{Potier2020ComparingFocalPlanea,Patterson2019DesignDescriptionCommissioning,Sidick2015StudiesEffectsControl}, the two that have been focusing on segmented apertures are the High-contrast imager for Complex Aperture Telescopes (HiCAT) testbed at the Space Telescope Science Institute \citep{Soummer2018HighcontrastImagerComplex} and the High Contrast Spectroscopy Testbed for Segmented Telescopes (HCST) at Caltech \citep{Llop-Sayson2020HighcontrastDemonstrationApodized}.

The tolerancing problem of segmented high contrast instruments has been previously addressed with the Pair-based Analytical model for Segmented Telescope Imaging from Space \citep[PASTIS;][]{Laginja2021AnalyticalTolerancingSegmented,Laginja2020PredictingContrastSensitivity,Laginja2019WavefrontErrorTolerancing,Leboulleux2018PairbasedAnalyticalModel,Leboulleux2018SensitivityAnalysisHighcontrast,Leboulleux2017SensitivityAnalysisHighcontrast}. It is an analytical model that calculates the average contrast in the DH caused by pupil-plane segment misalignments, using a simple matrix multiplication. Central to this model is the PASTIS matrix $M$, which describes the contrast contributions of an aberrated segment pair. When we combine it with a covariance matrix $C_a$ that describes the thermo-mechanical behavior of the segments, the PASTIS model can be used to calculate the expected mean contrast and its variance for a particular instrument, over many segmented aberration states, with analytical equations. This allows us to fully describe the statistical response of a segmented coronagraph to segment-level cophasing errors. Additionally, an eigendecomposition of either matrix allows us to write the contrast as a sum of separate contributions and then to invert the problem: instead of calculating the expected mean contrast given some aberrations, we can now statistically determine tolerances that lead to a particular mean contrast target. This leads to a quantitative tool for instrument design that provides means to calculate statistical limits on the segmented aberration modes.

In this paper, we use the semi-analytical development of the PASTIS model \citep{Laginja2021AnalyticalTolerancingSegmented,Laginja2019WavefrontErrorTolerancing} to measure an experimental PASTIS matrix on the HiCAT testbed \citep{Soummer2018HighcontrastImagerComplex,Moriarty2018HighcontrastImagerComplex,Leboulleux2016HighcontrastImagerComplex,Leboulleux2017ComparisonWavefrontControl,N'Diaye2015HighcontrastImagerComplex,N'Diaye2014HighcontrastImagerComplex,N'Diaye2013HighcontrastImagerComplex} by replacing the images usually provided by an end-to-end simulator with laboratory measurements. We then use this experimental PASTIS matrix for further analysis, meaning the validation of the instantaneous forward model for the calculation of the average DH contrast and WFE tolerancing analysis, and compare the results to a simulated case. First, we aim to provide a general experimental validation of the PASTIS model, showing that it is feasible to measure a PASTIS matrix on hardware, and that we can calculate correct segment-level tolerances in the mid-contrast regime ($10^{-6} - 10^{-8}$), compatible with the HiCAT performance as of today. This is demonstrated in Sect.~\ref{sec:results-hicat_exp_paper} at $2.5 \times 10^{-8}$, which is the contrast floor considered in this paper. Our tolerancing process is demonstrated for this value of contrast, situated at an intermediate point between the expectations of the James Webb Space Telescope (JWST) and LUVOIR. In Sect.~\ref{sec:results-hicat_exp_paper}, it translates into tolerancing values of a few nanometers, compatible with the hardware limitations of the bench (resolution of deformable mirrors, fast fluctuations in the optical system). We eventually demonstrate the sensitivity to a contrast change around a few $10^{-7}$. Second, we compare the results obtained with testbed measurements to results obtained with a purely simulated PASTIS matrix, and we analyze the sensitivity to model errors. And third, we establish an error budget for the segmented deformable mirror (DM) on HiCAT, for various contrast levels, and present quantitative results on the required WFE stability of the segmented mirror.

In Sect.~\ref{sec:recall-pastis_exp_paper} we recall the most important points about the PASTIS forward model and its inversion, from which we derive the results for WFE tolerancing. We also include a simple extension for the treatment of a drifting coronagraph floor that is not attributed to the segmented mirror. In Sect.~\ref{sec:hicat-testbed_exp_paper} we describe the HiCAT project and the testbed configuration used for the presented experiments. In Sect.~\ref{sec:results-hicat_exp_paper} we show the tolerancing results and their validations performed on the HiCAT testbed, and compare them to a simulated case of the PASTIS matrix on HiCAT. Finally, in Sects.~\ref{sec:discussion_exp_paper} and \ref{sec:CONCLUSIONS_exp_paper} we discuss our results and report our conclusions.

Note that the main figure of merit used by the PASTIS model is the spatially averaged intensity in the DH, normalized to the peak of the direct image, which is what we refer to as ``contrast" throughout this paper; it depends on the particular state of the instrument, and in particular of the segments. We also stress that we differentiate between this spatially averaged DH intensity, the ``average DH contrast", and a statistical mean (expected value) of this averaged contrast over many optical propagations, the statistical ``mean contrast".


\section{PASTIS tolerancing model and extension to a drifting coronagraph floor}
\label{sec:recall-pastis_exp_paper}

While segmented aberrations have a direct impact on the focal plane response, they are not the only source of time fluctuations in the contrast. For a laboratory validation, the environmental conditions (humidity, temperature, vibrations) evolve with time and contribute to opto-mechanical deformations of the testbed, eventually translating into slowly-evolving optical aberrations and contrast drift. We need to take this contrast drift, which is not due to the segmented mirror, into account to be able to isolate the effects coming from the segmented mirror alone.

In the following section, we present a brief summary of the PASTIS tolerancing model \citep{Laginja2021AnalyticalTolerancingSegmented}, and expand it to include a drifting coronagraph floor arising from time-dependent aberrations from sources other than the segmented mirror.

We model the phase $\phi$ in the pupil plane of a segmented high contrast instrument as:
    \begin{equation}
    \phi (\mathbf{r}, t) = \phi_{DH} (\mathbf{r}) + \phi_{ab} (\mathbf{r}, t) + \phi_{s} (\mathbf{r}),
    \label{eq:total-phase-composition_exp-paper}
    \end{equation}
where $\phi_{DH}$ is a static best-contrast phase solution, usually produced by a DH algorithm and applied to a pair of DMs. The term $\phi_{ab}$ is the phase produced by time-dependent aberrations in the system, $\phi_{s}$ is the phase caused by segment-induced aberrations, $\mathbf{r}$ is the pupil plane coordinate and $t$ the time variable. Under the assumption of the small aberration regime for $\phi_s$ and $\phi_{ab}$, and assuming that $\phi_{DH}$ is static, we discard any cross-terms between $\phi_s$ and $\phi_{ab}$ as they would create third and fourth order terms in contrast, while we limit ourselves to a second order model. We can then express the electric field in the pupil with two terms: the first is a time-dependent term that includes the best-contrast phase solution with an additional aberrating phase drift, and the second is independent from time and contains the segmented perturbations:
    \begin{equation}
    E (\mathbf{r}, t) = P(\mathbf{r})\,e^{i\,\phi (\mathbf{r},t)} \simeq P'(\mathbf{r},t) + i\,P(\mathbf{r})\,e^{i\,\phi_{DH}(\mathbf{r})} \phi_s(\mathbf{r}),
    \label{eq:phase-two-term}
    \end{equation}
where $P$ is the pupil function, and $P'(\mathbf{r},t) = P(\mathbf{r})\,\exp[i\, (\phi_{DH} (\mathbf{r}) + \phi_{ab} (\mathbf{r}, t))]$. Applying a linear coronagraph operator, $\mathcal{C}$, that represents the propagation of the electric field in the high contrast system (i.e., Fourier transforms and mask multiplications) to the expression given in Eq.~(\ref{eq:phase-two-term}), we obtain the coronagraphic intensity distribution in the image plane with $| \mathcal{C}\{E (\mathbf{r}, t)\} |^2$. The average contrast is then given by averaging over the DH area, indicated by $\langle \dots \rangle_{DH}$.

It was previously shown that the average DH contrast can always be expressed as a quadratic function of a segmented phase perturbation under an appropriate change of variable \citep[Eqs.~4 and 11]{Laginja2021AnalyticalTolerancingSegmented}, which eliminates the linear cross-term, and leaves us with separate square transformations of the two terms in Eq.~(\ref{eq:phase-two-term}). Concretely, in this paper we model the contrast floor with a contribution from the static DM phase, and aberrations introduced by environmental changes of the testbed, which cause a drift in the contrast as a function of time $t$:
    \begin{equation}
    c_0(t)  = \langle |\mathcal{C}\{ P'(\mathbf{r},t)\}|^2\rangle_{DH} = \langle |\mathcal{C}\{P(\mathbf{r}) e^{i\,(\phi_{DH}(\mathbf{r}) + \phi_{ab}(\mathbf{r}, t))}\}|^2\rangle_{DH}.
    \label{eq:c0-clc-wfsc_exp-paper}
    \end{equation}

Representing optical aberrations on a segmented telescope with local (per-segment) Zernike modes, we can expand the phase aberrations on the segmented pupil $\phi_s (\mathbf{r})$ in the second term of Eq.~(\ref{eq:phase-two-term}) as a sum of segment-level polynomials  \citep[Eq.~1]{Laginja2021AnalyticalTolerancingSegmented}, with its decomposition on such a basis denoted as $\mathbf{a}$. Following the development of the original PASTIS model, we can express the average contrast in the coronagraphic DH as a matrix multiplication \citep[Eq.~9]{Laginja2021AnalyticalTolerancingSegmented}, which makes our assumed model:
    \begin{equation}
    c(t) = c_0(t) + \mathbf{a}^T M \mathbf{a},
    \label{eq:pastis-equation_exp-paper}
    \end{equation}
where $c(t)$ is the spatial average contrast in the DH, $c_0(t)$ the coronagraph floor (i.e., the average contrast in the DH in the presence of the best-contrast phase $\phi_{DH}(\mathbf{r})$ and of the variable phase aberrations $\phi_{ab} (\mathbf{r}, t)$, but in the absence of segment misalignments $\phi_s$), $M$ is the symmetric PASTIS matrix of dimensions $n_{seg} \times n_{seg}$ with elements $m_{ij}$, $\mathbf{a}$ is the aberration vector of the local Zernike coefficients on all discrete $n_{seg}$ segments and $\mathbf{a}^T$ its transpose. We can see that the contrast floor $c_0(t)$ is dominated by the DH phase solution $\phi_{DH} (\mathbf{r})$, with an additional variation introduced by the time-dependent $\phi_{ab} (\mathbf{r}, t)$. The matrix $M$ itself contains a constant term added by $e^{i\,\phi_{DH}(\mathbf{r})}$, and the segmented aberrations induced by $\phi_s (\mathbf{r})$. Following the pair-wise aberrated approach explained previously \citep[Eq.~10]{Laginja2021AnalyticalTolerancingSegmented}, the matrix elements can thus generally be expressed as:
    \begin{equation}
    m_{ij} = \langle \mathcal{C}\{P(\mathbf{r})\,e^{i \phi_{DH}(\mathbf{r})} Z(\boldsymbol{\mathbf{r}}-\mathbf{r_i}) \} \mathcal{C}\{P(\mathbf{r})\,e^{i \phi_{DH}(\mathbf{r})} Z(\boldsymbol{\mathbf{r}}-\mathbf{r_j}) \}^* \rangle_{DH}.
    \label{eq:matrix-elements_exp-paper}
    \end{equation}
By defining the differential contrast as our objective quantity that is independent of time $t$:
    \begin{equation}
    \Delta c = c(t) - c_0(t),
    \label{eq:delta-c-objective_exp-paper}
    \end{equation}
we render the right-hand side of Eq.~(\ref{eq:pastis-equation_exp-paper}) time-independent, which allows us to isolate the effects imposed by segment cophasing errors, defined by the vector $\mathbf{a}$.

Each PASTIS matrix element $m_{ij}$ represents the contrast contribution to the DH average contrast $c_{ij}$ by each aberrated segment pair in the pupil, formed by segments $i$ and $j$. Once the matrix is established, we can calculate its eigenmodes $\mathbf{u}_p$ and eigenvalues $\lambda_p$ by means of an eigendecomposition. The total number of optical (PASTIS) modes, $n_{modes}$, is equal to the total number of segments, $n_{seg}$. Since the eigenmodes are orthonormal and diagonalize $M$, the DH contrast can be written as the sum of separate contributions of each mode, and each eigenvalue is the contrast sensitivity of the corresponding mode \citep[Eq.~22]{Laginja2021AnalyticalTolerancingSegmented}:
    \begin{equation}
    \Delta c = \sum_{p}^{n_{modes}} b_p^2 \lambda_p,
    \label{eq:get-bs-from-matrices_exp-paper}
    \end{equation}
where $b_p$ is the amplitude of the $p^{th}$ mode.

With the PASTIS matrix $M$ representing the optical properties of the segmented coronagraph, and Eq.~(\ref{eq:pastis-equation_exp-paper}) giving the instantaneous average DH contrast c for a given aberration vector \textbf{a}, $M$ can be combined with any given segment covariance matrix $C_a$ to calculate the statistical mean and variance of the average DH contrast  \citep[Eqs.~31 and 32]{Laginja2021AnalyticalTolerancingSegmented}:
    \begin{equation}
    \langle \Delta c \rangle  = \tr(M C_a),
    \label{eq:mean-delta-c_exp-paper}
    \end{equation}
where $\tr$ denotes a trace, and:
    \begin{equation}
    \Var (\Delta c) = 2 \tr [(M C_a)^2].
    \label{eq:var-of-c_exp-paper}
    \end{equation}
While Eq.~(\ref{eq:mean-delta-c_exp-paper}) does not make any assumptions about the statistics of the vector $\mathbf{a}$, Eq.~(\ref{eq:var-of-c_exp-paper}) is true when $\mathbf{a}$ follows a Gaussian distribution, which is an assumption used throughout this paper. The two equations above allow us to calculate these two integral quantities directly from the optical properties of the instrument, described by $M$, and the mechanical correlations of the segments, captured by $C_a$, which can be obtained from thermo-mechanical modeling of the observatory.

For all $C_a$, Eq.~(\ref{eq:mean-delta-c_exp-paper}) can be expressed as
    \begin{equation}
    \langle\Delta c\rangle = \sum_{p=1}^{n_{modes}} \sigma_{b_p}^2 \lambda_p,
    \label{eq:delta-c-from-bps_exp-paper}
    \end{equation}
where $\sigma_{b_p}$ are the standard deviations of the optical mode amplitudes, in the diagonalized basis of the PASTIS matrix $M$. This leads to an inversion of the problem where we set a differential target contrast $\Delta c$, for which we want to derive WFE tolerancing limits in terms of standard deviations for the segments, or modes. For the special case of a diagonal $C_a$, which means that the individual segments are statistically independent, a similar expression to Eq.~(\ref{eq:delta-c-from-bps_exp-paper}) can be deduced from Eq.~(\ref{eq:mean-delta-c_exp-paper}) for the standard deviations of the segment amplitudes, $\sigma_{a_k}$:
    \begin{equation}
    \langle\Delta c\rangle = \sum_{k=1}^{n_{seg}} \sigma_{a_k}^2 m_{kk},
    \label{eq:get-muk-from-matrices_exp-paper}
    \end{equation}
where the $m_{kk}$ are the diagonal elements of the PASTIS matrix. This equation can be used to specify the standard deviation for each segment: denoting $\mu_k = \sigma_{a_k}$, we can choose that every segment contributes equally to the contrast, which yields a specification of segment amplitude standard deviations of \citep[Eq.~36]{Laginja2021AnalyticalTolerancingSegmented}:
    \begin{equation}
    \mu_k^2  = \frac{\langle\Delta c\rangle }{n_{seg} m_{kk}}.
    \label{eq:mus-delta-c_exp-paper}
    \end{equation}

While Eqs.~(\ref{eq:mean-delta-c_exp-paper}) and (\ref{eq:var-of-c_exp-paper}) allow us to analytically calculate the expected mean contrast of a segmented coronagraph and its variability, for mechanical properties described by $C_a$, Eq.~(\ref{eq:mus-delta-c_exp-paper}) provides a way to determine individual segment tolerances for a particular target differential contrast $\Delta c$ that is to be maintained over a set of observations.


\section{The HiCAT project and experimental setup}
\label{sec:hicat-testbed_exp_paper}

The High-contrast imager for Complex Aperture Telescopes testbed \citep[HiCAT;][]{Soummer2018HighcontrastImagerComplex,Moriarty2018HighcontrastImagerComplex,Leboulleux2016HighcontrastImagerComplex,Leboulleux2017ComparisonWavefrontControl,N'Diaye2015HighcontrastImagerComplex,N'Diaye2014HighcontrastImagerComplex,N'Diaye2013HighcontrastImagerComplex} is  dedicated to a LUVOIR-type coronagraphic demonstration with on-axis segmented apertures\footnote{\url{https://exoplanets.nasa.gov/internal_resources/1186/}}. The project is targeting experiments in ambient conditions that can happen before demonstrations in a vacuum, for example at the Decadal Survey Testbed \citep[DST;][]{Patterson2019DesignDescriptionCommissioning} located at the Jet Propulsion Laboratory. The ultimate performance goal of such testbeds is to demonstrate a contrast of $10^{-10}$ in the lab. While this goal can only be achieved in an environmentally stable vacuum chamber, the HiCAT testbed is aiming for $10^{-8}$, limited by its coronagraph performance and environmental conditions. The work on HiCAT intends to provide a system-level analysis of a high contrast instrument that includes various sensors and controllers. Ultimately, the planned coronagraph for HiCAT operations is an Apodized Pupil Lyot Coronagraph \citep[APLC;][]{N'Diaye2016APODIZEDPUPILLYOT,Zimmerman2016LyotCoronagraphDesign,N'Diaye2015ApodizedPupilLyot} that includes apodizers manufactured using carbon nanotubes. Since the various apodizer designs are mounted on easily interchangeable bonding cells, a high-quality flat mirror can be swapped in to use a classical Lyot coronagraph (CLC). We use the CLC setup (Sect.~\ref{subsec:hicat-for-pastis_exp_paper}) for the experiments in this paper, supported by an active WFS\&C loop to improve the DH contrast beyond the initial static solution caused purely by the coronagraphic masks (Sect.~\ref{subsec:active-wfsc_exp_paper}). An IrisAO \citep{Helmbrecht2013HighactuatorcountMEMSDeformable} segmented DM is utilized as the segmented telescope simulator on the testbed to introduce segment-level WFE for the tolerancing validation with PASTIS.

\subsection{Classical Lyot coronagraph as static setup}
\label{subsec:hicat-for-pastis_exp_paper}

While the HiCAT APLC is designed to provide a superior performance on a segmented aperture compared to the simpler CLC, the pupil plane apodization of this coronagraph causes a lot of the aperture segments to be highly concealed \citep[Fig.~7]{Soummer2018HighcontrastImagerComplex}.  HiCAT has been operated as a segmented CLC since the fall of 2020, and we used this testbed configuration to perform the experimental validations of PASTIS, which allows us to image the entire segmented aperture (unobstructed 37 segments).

\begin{figure}
   \resizebox{\hsize}{!}{\includegraphics{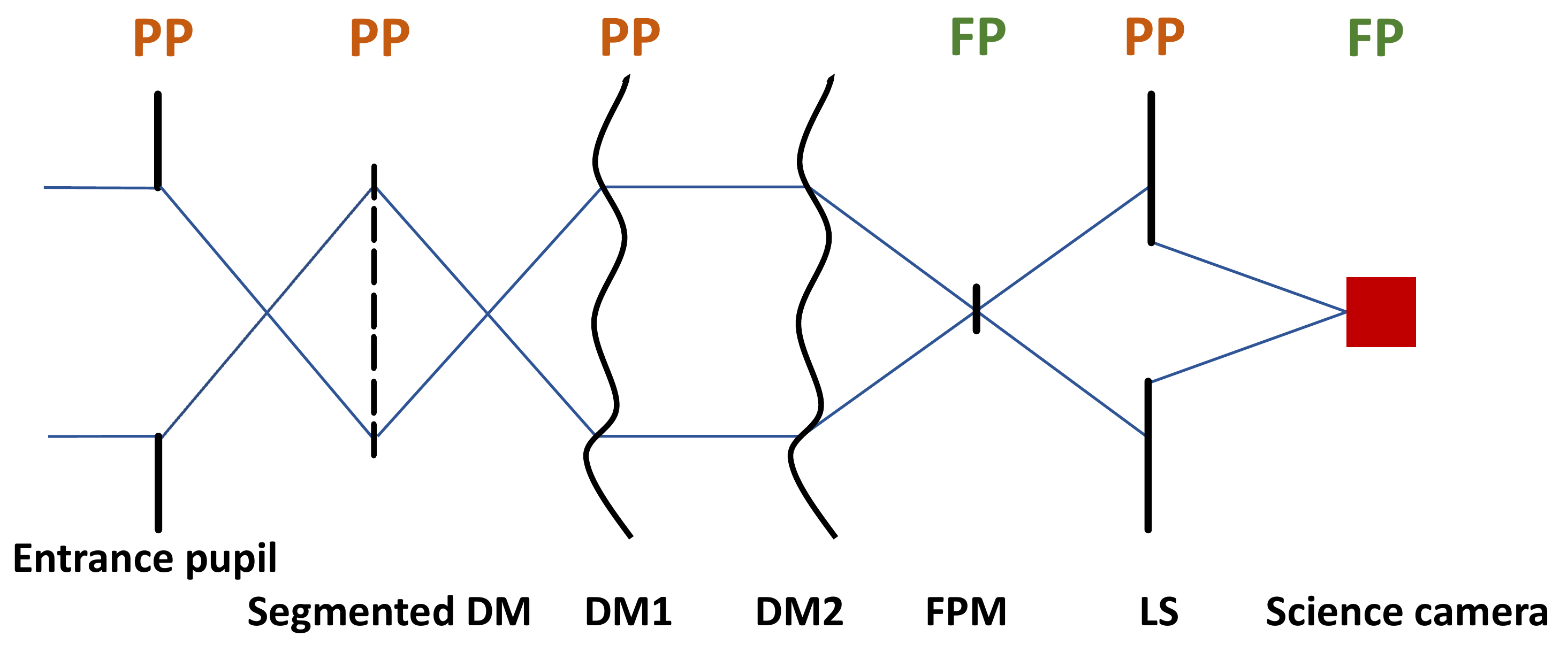}}
   \caption[HiCAT layout for PASTIS] 
   {\label{fig:hicat_layout_pastis_experimental} 
    HiCAT optical configuration used for PASTIS experiments, here shown with transmissive optics for simplicity. The entrance pupil is a custom shaped mask tracing the outline of the segmented DM in a consecutive pupil plane, see also Fig.~\ref{fig:hicat_pupil_overlaps_experimental}. Of the two continuous deformable mirrors, DM1 is in a pupil plane, and DM2 is located out of pupil in order to control both phase and amplitude. The focal plane mask (FPM) and Lyot stop (LS) build the classical Lyot coronagraph (CLC) setup. The pupil planes (PP) and focal planes (FP) are marked.}
   \end{figure}

\begin{figure}
    \resizebox{\hsize}{!}{\includegraphics{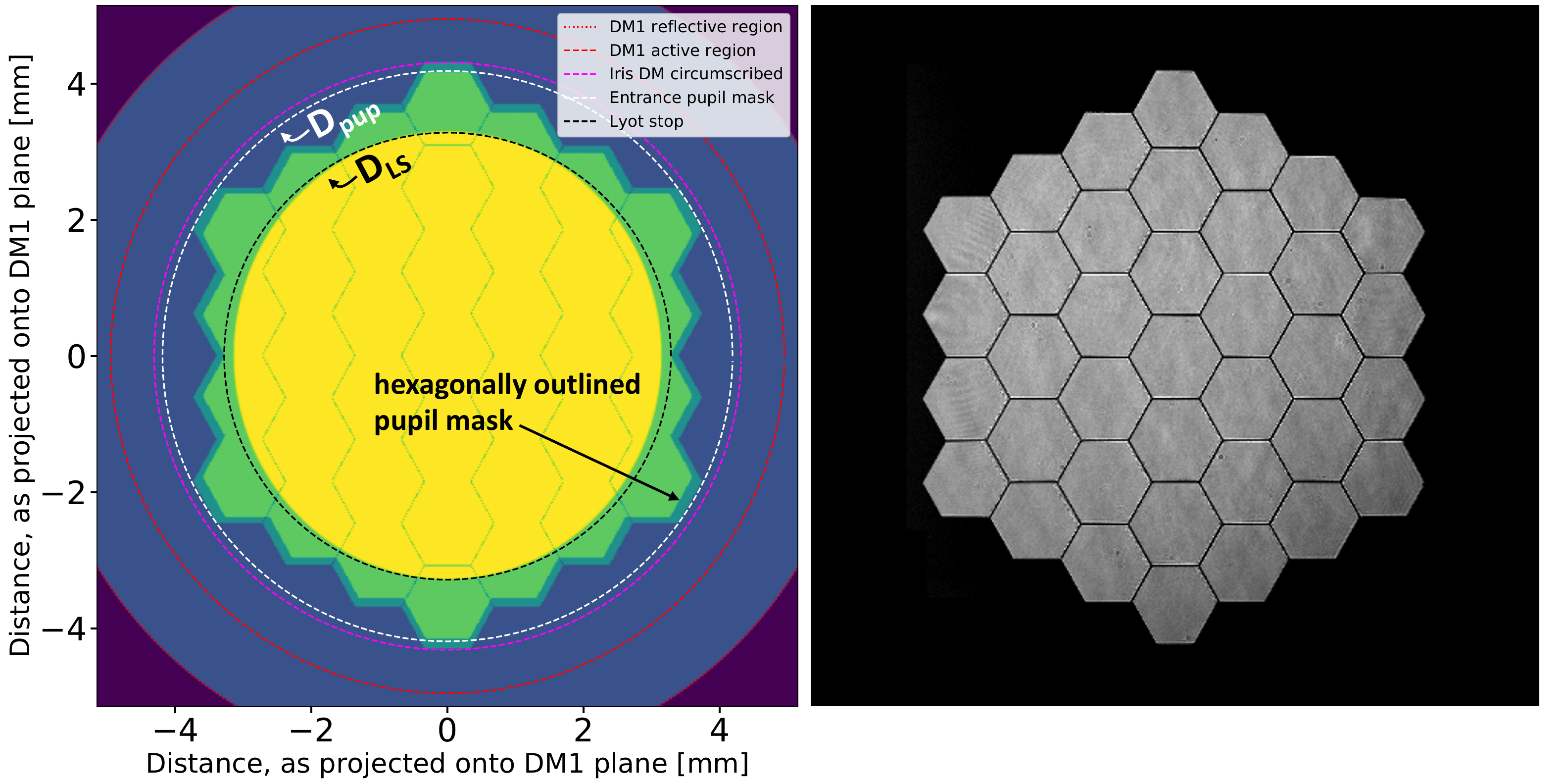}}
   \caption[HiCAT pupil overlaps and hardware pupil image] 
   {\label{fig:hicat_pupil_overlaps_experimental} 
    \textit{Left:} Overlapping pupils in the HiCAT segmented CLC configuration, and their diameters, used for the PASTIS experiments, projected in the pupil plane of DM1 (simulated image). The entrance pupil mask (bright green, undersized polygon shape) traces the outline of the IrisAO (dark green polygon shape) with a 97\% undersizing factor, preventing the illumination of areas outside of the controllable segments. The entrance pupil diameter $D_{pup}$ (white dashed line) is defined as the circumscribed circle around the undersized pupil mask. The LS (yellow ellipse) is sized such that its edges stay within the controllable outline of the IrisAO. Since HiCAT uses reflective optics, the resulting beam foreshortening along the x-axis results in all pupils being optically ``squished" along the x direction. This is true for all optics and is immediately visible in this figure as the yellow LS surface is slightly elliptical with respect to the circle denoting its nominal diameter $D_{LS}$ (black dashed line). \textit{Right:} Measured pupil image on hardware with a detector located in a pupil plane before the two continuous DMs and the LS, showing the IrisAO segments and the pupil mask outlining the segmented DM. Note the slightly undersized outline, which results in somewhat irregular hexagons at the edges, especially noticeable on the six corner segments.}
   \end{figure}

The defining optical elements of HiCAT with a CLC are a non-circular pupil mask, an IrisAO PTT111L 37-element hexagonally-segmented DM \citep{Helmbrecht2016LongtermStabilityTemperature,Helmbrecht2013HighactuatorcountMEMSDeformable}, two Boston Micromachines 952-actuator microelectro-mechanical (MEMS) “kilo-DMs” \citep{Cornelissen2010MEMSDeformableMirrors}, a focal plane mask (FPM), a Lyot stop (LS) and a science detector. A schematic of the optical testbed layout used for the presented experiments can be seen in Fig.~\ref{fig:hicat_layout_pastis_experimental}.
The pupil mask traces the hexagonally segmented IrisAO outline, and we use its circumscribed diameter as the pupil diameter, $D_{pup}$. This is slightly undersized with respect to the diameter of the IrisAO itself to limit the beam to the controllable area of the segmented DM. The first Boston continuous DM is located in a pupil plane, and the second one is out-of-pupil, at a distance of 30~cm from the pupil plane DM, in order to control amplitude in the WFS\&C process. The FPM has a diameter of $8.52\, \lambda/D_{pup}$, with $\lambda$ (no subscript) the central wavelength of the bandpass. The LS is a circular, unobscured mask with a diameter $D_{LS}$ of 79\% of the size of $D_{pup}$, as projected in the Lyot plane. An overlay of all relevant pupils can be seen in Fig.~\ref{fig:hicat_pupil_overlaps_experimental}.

Previously, the IrisAO had been used with the CLC for experiments on coronagraphic focal plane wavefront sensing (WFS) on a segmented aperture \citep{Leboulleux2020ExperimentalValidationCoronagraphic}, but with different mask sizes and a fully circular entrance pupil. The installation of the IrisAO on the current CLC setup was performed in late 2020. The segments of the IrisAO segmented DM are each controllable in piston, tip and tilt, with a maximum stroke of $5~\mu$m on each of the three actuators mounted on the back side of each segment. The segmented DM initially saw an open-loop flatmap calibration \citep{Helmbrecht2016LongtermStabilityTemperature} with a 4D Fizeau interferometer, which yielded a calibrated surface error of 9~nm root-mean-square (rms) \citep[Fig.~3]{Soummer2018HighcontrastImagerComplex}. This was improved upon after installing the IrisAO on the testbed, where a finer, closed-loop flatmap calibration was performed with dOTF phase retrieval \citep{Codona2012TheoryApplicationDifferential, Codona2012ExperimentalEvaluationDifferential}.

The average contrast for the currently used CLC setup, in an annular DH from 6-10 $\lambda/D_{LS}$ with flattened DMs is ${\sim} 1 \times 10^{-5}$ in monochromatic light at 638~nm. In order to place the coronagraph floor of the testbed into a higher contrast regime, we deploy an iterative WFS\&C loop described in the following section.

\subsection{Active wavefront sensing and control for an improved DH}
\label{subsec:active-wfsc_exp_paper}

\begin{figure*}
   \centering
   \includegraphics[width = 17cm]{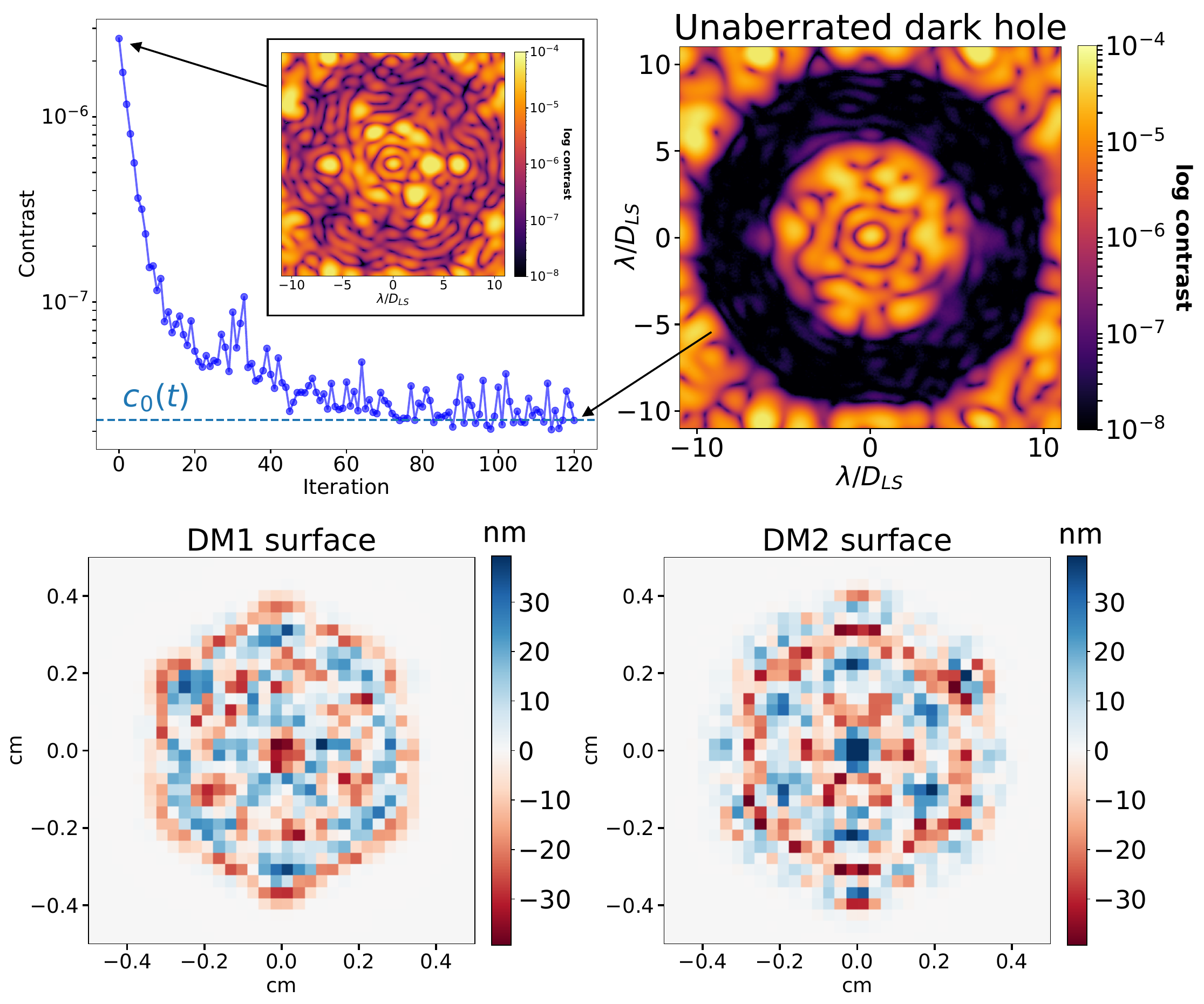}
   \caption[Baseline DH] 
   {\label{fig:dms-and-dh_experimental} 
    \textit{Top:} 120 iterations of pair-wise sensing and stroke minimization (left), starting from a pre-modulated coronagraphic image shown in the embedded plot. The loop converges after 70-80 iterations and yields an average contrast of $2.5 \times 10^{-8}$ from 6-10 $\lambda/D_{LS}$, using a monochromatic source at 638~nm (right). The two DH images are shown on the same scale. \textit{Bottom:} DM surface commands applied to continuous DM1 (in-pupil, left) and DM2 (out-of-pupil, right) for the best-contrast DH solution in the top right at iteration 120. The segmented DM in this setup is statically set to its best flat position throughout the control loop.}
   \end{figure*}

The WFS\&C strategy used on HiCAT to improve the monochromatic DH contrast in an annular DH deploys an iterative approach of pair-wise probing \citep{Groff2016MethodsLimitationsFocal,Give'on2011PairwiseDeformableMirror} to estimate the electric field, and stroke minimization \citep{Mazoyer2018a:ActiveCorrectionApertureI,Mazoyer2018b:ActiveCorrectionApertureII,Pueyo2009OptimalDarkHole} for control. The outer working angle of 10 $\lambda/D_{LS}$ is defined by the highest spatial frequency controllable by the two continuous DMs, and the IrisAO DM is kept at its best flat position throughout. In order to avoid a local minimum, we first dig a larger DH at moderate contrast before launching the loop on a 6--10 $\lambda/D_{LS}$ DH as seen in Fig.~\ref{fig:dms-and-dh_experimental}. After 70--80 iterations, the contrast performance converges to $2.5 \times 10^{-8}$, with variations on the order of $2 \times 10^{-8}$ during the WFS\&C sequence. The continuous DM commands that create the final DH, as well as the convergence plot and the final DH image, are shown in Fig.~\ref{fig:dms-and-dh_experimental}.
The DM surface commands shown in Fig.~\ref{fig:dms-and-dh_experimental} are applied at the beginning of each experiment presented in Sect.~\ref{sec:results-hicat_exp_paper}, making them a part of the static coronagraph contribution $c_0(t)$ as described by Eq.~(\ref{eq:c0-clc-wfsc_exp-paper}). This setup sets our nominal coronagraph floor that we use in the PASTIS experiments on HiCAT in Sect.~\ref{sec:results-hicat_exp_paper} to an initial $2.5 \times 10^{-8}$, and it is drifting without active control during the experiments. The aberrations we target with the PASTIS tolerancing model in this paper are the segmented WFE, $\mathbf{a}$, introduced by the IrisAO DM on top of this static best-contrast solution, and independent of the contrast drift.


\section{Experimental validation of segmented tolerancing on the HiCAT testbed}
\label{sec:results-hicat_exp_paper}

In the following section, we present the results of several experiments for the hardware validation of the PASTIS tolerancing model. We measure an experimental PASTIS matrix and compare it to simulations, and we confirm the instantaneous PASTIS forward model in Eq.~(\ref{eq:pastis-equation_exp-paper}). We measure the deterministic optical mode contrast given by Eq.~(\ref{eq:get-bs-from-matrices_exp-paper}) and then validate the statistical segment tolerances calculated from the experimental PASTIS matrix with Eq.~(\ref{eq:mus-delta-c_exp-paper}) by performing Monte Carlo experiments and comparing them to results from Eqs.~(\ref{eq:mean-delta-c_exp-paper}) and (\ref{eq:var-of-c_exp-paper}). We use the testbed configuration described in Sect.~\ref{sec:hicat-testbed_exp_paper}, and we describe how to correct our measurements for the drifting contrast floor $c_0(t)$ in Sect.~\ref{subsec:measured-pastis-matrix_exp_paper}. Before running an experiment, we apply the continuous DM solutions for the DH shown in Fig.~\ref{fig:dms-and-dh_experimental}, putting the testbed initially at $c_0(t_0) = 2.5 \times 10^{-8}$. In Sect.~\ref{subsec:measured-pastis-matrix_exp_paper}, we emphasize the need to refine our forward model to account for the slow contrast drift on the testbed, as introduced in Sect.~\ref{sec:recall-pastis_exp_paper}. At this level of performance, this drift is the main limitation and has to be corrected numerically. In Sect.~\ref{subsec:segment-tolerances_exp_paper}, we compute the tolerancing in terms of segment allocations corresponding to delta contrast values of a few $10^{-7}$, as limited by the uncorrected fast fluctuations we see in our data in Fig.~\ref{fig:contrast-drift-matrix-piston_experimental}.

\subsection{PASTIS matrix measurement and deterministic forward model validation}
\label{subsec:measured-pastis-matrix_exp_paper}

\subsubsection{Measurement method}
The PASTIS matrix is a pair-wise influence matrix, linking segment aberrations to the differential average contrast in the coronagraphic DH. The total number of intensity measurements needed for the construction of an experimental PASTIS matrix is:
    \begin{equation}
    n_{meas} = \frac{n_{seg}(n_{seg}+1)}{2}.
    \label{eq:number-of-measurements_exp-paper}
    \end{equation}
Indeed, the matrix being symmetrical, we measure only the non-repeating permutations of segment pairs including the matrix diagonal. On the 37-segment HiCAT pupil this requires $n_{meas} = 703$ measurements.

The relation between each pair-wise aberrated contrast measurement and the PASTIS matrix elements is given by \citep[Eq.~15]{Laginja2021AnalyticalTolerancingSegmented}:
    \begin{equation}
     c_{ij}(t) =  c_0(t) + a_i^2 m_{ii} + a_j^2 m_{jj} + 2 a_i a_j\ m_{ij},
     \label{eq:almost-there-off_exp-paper}
    \end{equation}
where $a_i$ is the WFE amplitude on segment $i$. We use the same calibration aberration amplitude $a_c = a_i = a_j$ that is put on each individual segment in the measurement of the contrast matrix ($c_{ij}$). The calibration of the M matrix is thus obtained by the measurement of the contrast from pushing a pair of segments ($i$,$j$). Since the natural testbed contrast is evolving with time, this measurement must be corrected for the coronagraph floor persisting at that same time, $c_0(t)$, which can be easily remeasured for each $c_{ij}(t)$. The expression for the diagonal matrix elements then becomes:
    \begin{equation}
     m_{ii} = \frac{c_{ii}(t) - c_0(t)}{a_c^2}.
     \label{eq:drifting-diagonal_exp-paper}
    \end{equation}
This makes the PASTIS matrix diagonal entirely independent of time and the coronagraph floor, and it describes the contrast contribution of each individual segment to the DH. Ideally, $c_{ii}(t)$ and $c_0(t)$ are measured at the same time $t$; in reality, they are measured within a short time of each other, which needs to be faster than the occurring drifts and can thus be assumed to be simultaneous. Since this corrects the diagonal matrix elements for the coronagraph floor at the time of their measurement, we now want to make the off-diagonal elements depend on the already calibrated diagonal PASTIS matrix elements $m_{ii}$, rather than the uncalibrated diagonal measurements of the contrast $c_{ii}$. We can easily solve Eq.~(\ref{eq:almost-there-off_exp-paper}) for the off-diagonal PASTIS matrix elements:
    \begin{equation}
     m_{ij} = \frac{c_{ij}(t) - c_0(t)}{2 a_c^2} - \frac{m_{ii} + m_{jj}}{2}.
     \label{eq:drifting-off-diagonal_exp-paper}
    \end{equation}
In this way, each matrix element $m_{ij}$ gets calibrated with an appropriate, time-dependent measurement of $c_0(t)$, which makes the entire PASTIS matrix time-independent.

\subsubsection{Matrix measurement}
We use this to measure an experimental PASTIS matrix on the HiCAT testbed. We constrain ourselves to a local piston mode with a WFE amplitude of $a_c=40$~nm rms for the calibration aberration of the PASTIS matrix. Other modes are possible, for example tip/tilt, or a combination of local segment aberrations, but they are not considered in this paper. An aberration of 40~nm rms on a single segment of a 37-segment pupil translates to a global aberration of 6.6~nm rms, while two such aberrated segments cause a global WFE of 9.3~nm rms. We can see in Fig.~\ref{fig:hockeystick_experimental} that this puts $a_c$ significantly off the knee around the coronagraph floor, which was discussed as an optimal regime for $a_c$ in \cite{Laginja2021AnalyticalTolerancingSegmented}. This was done in order to increase the signal-to-noise ratio (SNR) in the fringe images during the matrix calibration, while simultaneously not increasing $a_c$ too much along the linear aberration regime indicated in Fig.~\ref{fig:hockeystick_experimental}.

\begin{figure}
    \resizebox{\hsize}{!}{\includegraphics{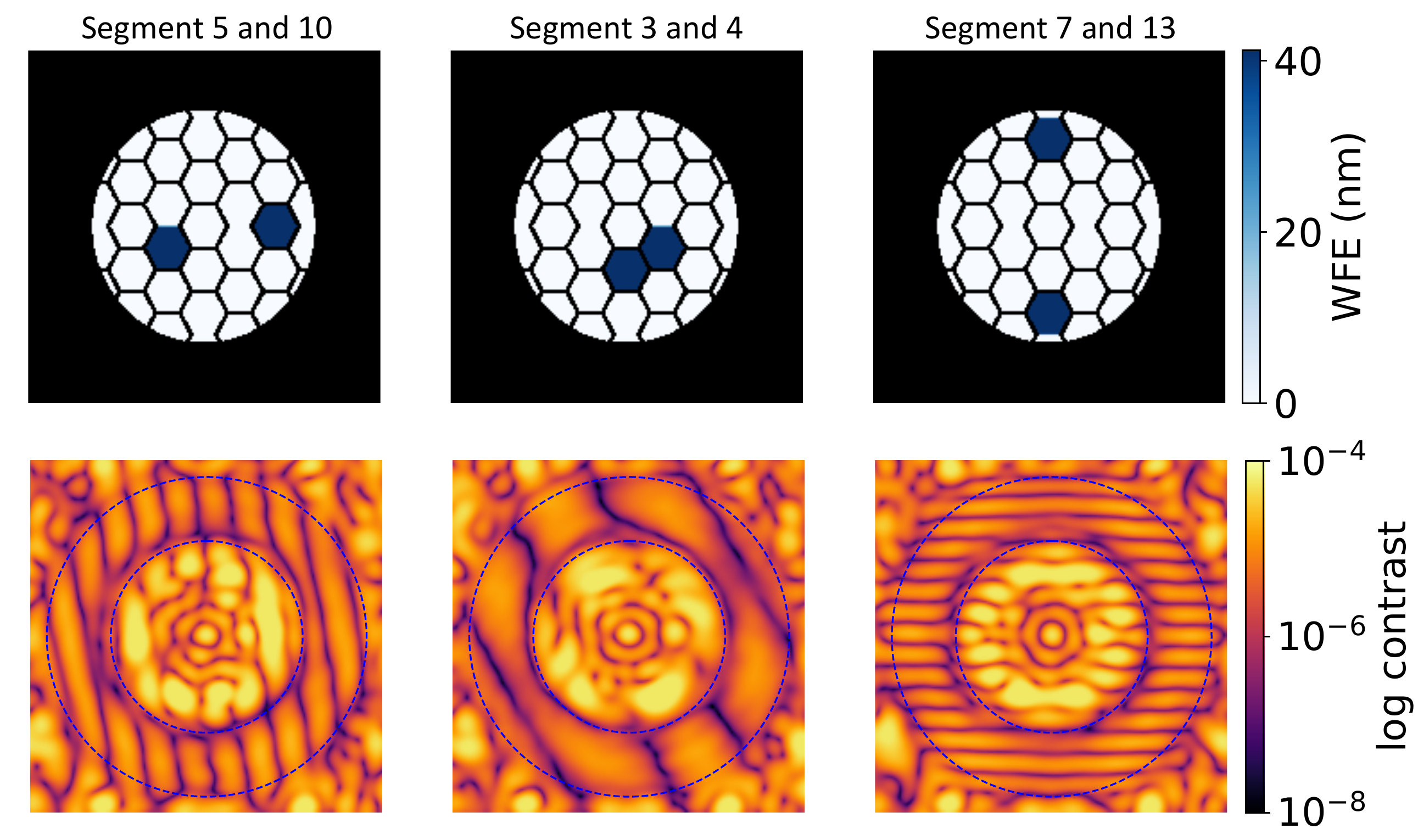}}
   \caption[Fringe images] 
   {\label{fig:fringe-images_experimental}
    Examples of pair-wise aberrated DH images. \textit{Top:} Simulated wavefront error maps of aberrated segment pairs on the IrisAO segmented DM. The LS cuts off most of the segments in the outer ring of the segmented pupil. Segment numbering as indicated in Fig.~\ref{fig:pastis-matrix_experimental}, left. \textit{Bottom:} DH images from the testbed with resulting fringes from pair-wise aberrated segments during the PASTIS matrix measurement shown in the top row. The DH extent is from 6-10 $\lambda/D_{LS}$, indicated with the dashed circles, and the images are displayed on the same range like the unaberrated DH in Fig.~\ref{fig:dms-and-dh_experimental}. The pair (3|4) is made of adjacent segments, which produces low-spatial frequency fringes in the DH. This leads to an overall decrease in the contrast contribution, as is confirmed by their respective entry in the PASTIS matrix in Fig.~\ref{fig:pastis-matrix_experimental}, with a blue entry right next to the matrix diagonal.}
   \end{figure}

\begin{figure}
    \resizebox{\hsize}{!}{\includegraphics{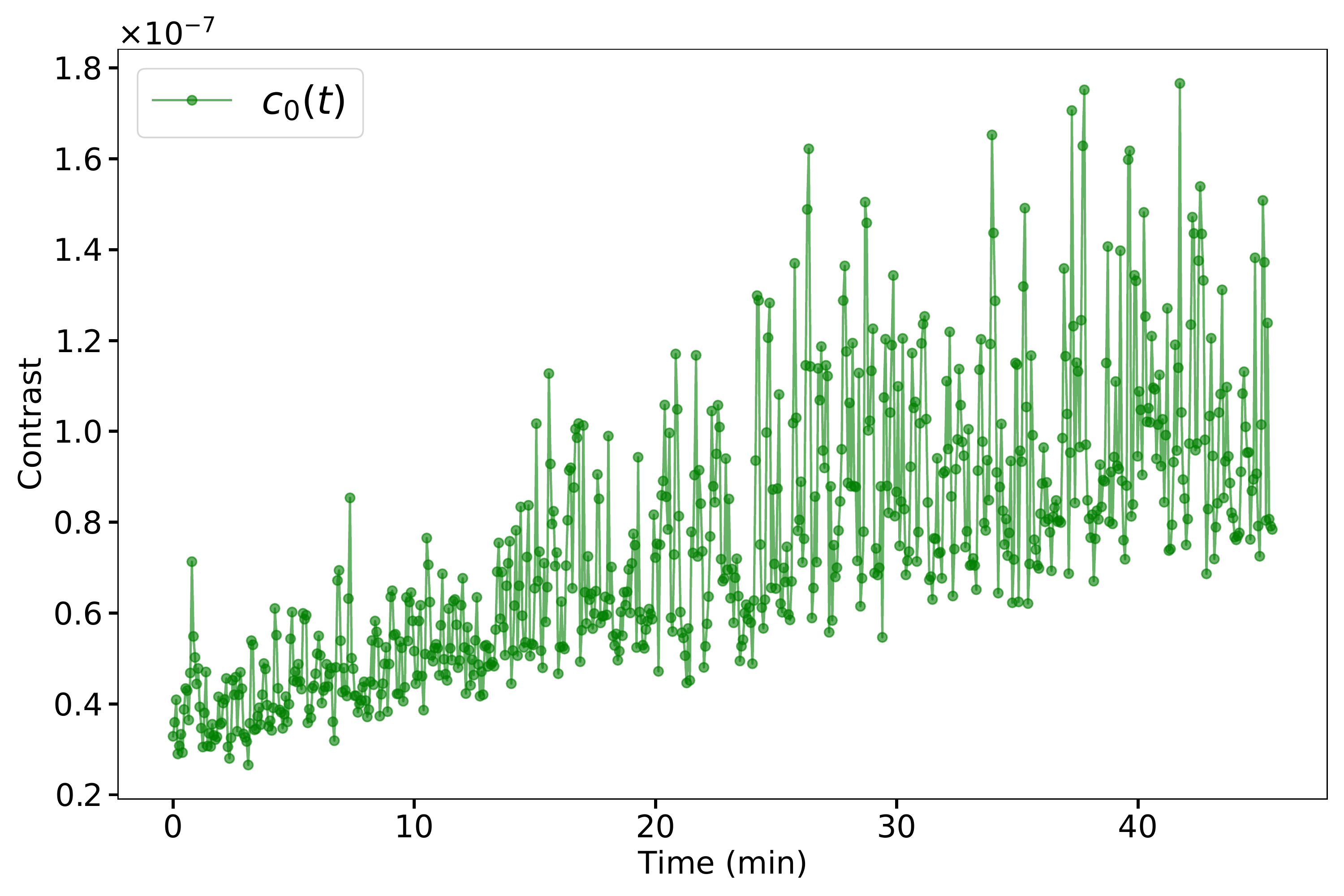}}
   \caption[Contrast drift piston matrix] 
   {\label{fig:contrast-drift-matrix-piston_experimental} 
    Contrast $c_0(t)$ during the PASTIS matrix acquisition. After each pair-wise aberrated DH measurement, we flatten the IrisAO segmented DM to measure the drift in the coronagraph floor over the course of the experiment. We subtract this off our data in order to perform an analysis on the differential contrast $\Delta c$. This open-loop contrast degrades gradually over time, note the linear scale. The measured contrast values range from $2.5 \times 10^{-8}$ to $8 \times 10^{-8}$ during the course of the experiment, but the difference between adjacent measurements is initially on the order of $2 \times 10^{-8}$, rising to $1 \times 10^{-7}$ later on, which is sufficient for our proposed calibration method. The total duration of the experiment is 45 minutes.}
   \end{figure}

\begin{figure*}
   \centering
    \includegraphics[width = 17cm]{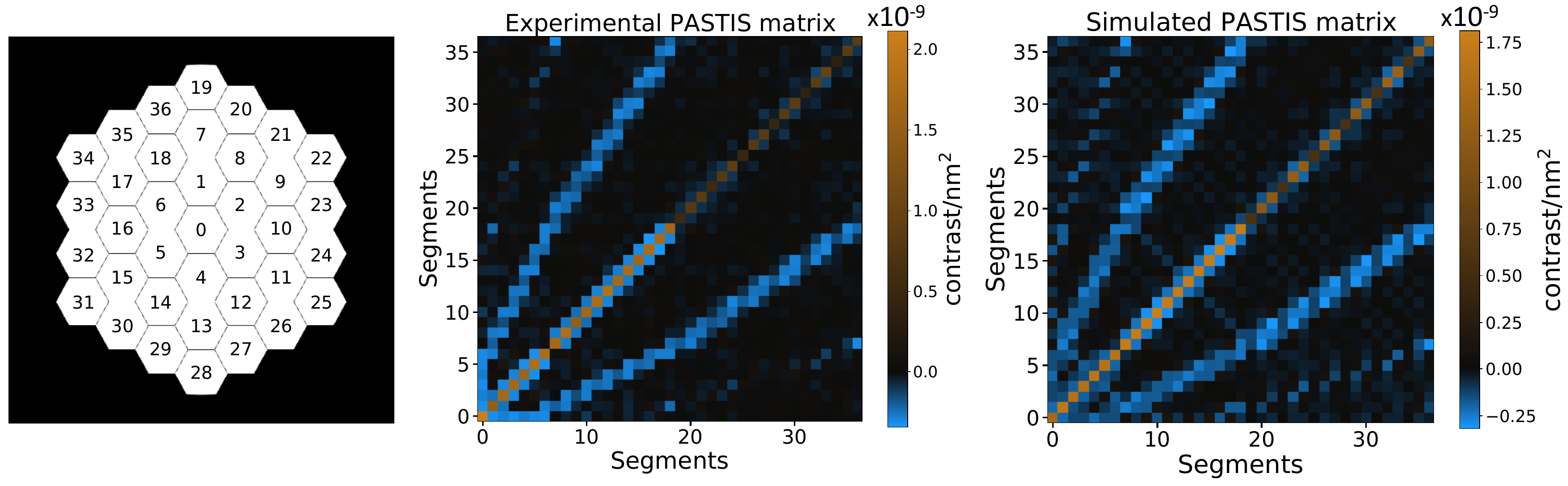}
   \caption[PASTIS matrix] 
   {\label{fig:pastis-matrix_experimental} 
    \textit{Left:} Geometry of the IrisAO segmented DM on HiCAT and the segment numbering used in this paper, in the entrance pupil. The 37 segments are numbered starting at 0 for the center segment, to 36 in the outer ring. In the exit pupil, most of the outer ring segments are obscured by the LS, see Fig.~\ref{fig:fringe-images_experimental}. \textit{Middle:} Experimental PASTIS matrix for HiCAT as measured on the testbed. Each entry represents the differential contrast contribution of each aberrated segment pair. The matrix is symmetric, and its diagonal shows the impact on the contrast by the individual segments. \textit{Right:} Simulated PASTIS matrix for HiCAT, without any WFE or noise in the optical system. This matrix shows the idealized contrast contributions from each segment pair in a perfect system.}
   \end{figure*}

We sequentially aberrate pairs of segments by applying the calibration aberration $a_c$ to each segment to measure $c_{ij}(t)$, and then flatten the IrisAO DM and record the coronagraph floor $c_0(t)$ for the same iteration. Examples of pair-wise aberrated DHs causing fringe patterns are displayed in Fig.~\ref{fig:fringe-images_experimental}, and the evolution of the unaberrated coronagraph floor during the PASTIS matrix acquisition is shown in Fig.~\ref{fig:contrast-drift-matrix-piston_experimental}.
We then use Eqs.~(\ref{eq:drifting-diagonal_exp-paper}) and (\ref{eq:drifting-off-diagonal_exp-paper}) to calculate the elements $m_{ij}$ and construct the experimental PASTIS matrix $M^{\text{exp}}$ shown in Fig.~\ref{fig:pastis-matrix_experimental}, middle. The PASTIS matrix is symmetric, with its diagonal describing the impact on the contrast by the individual segments. There are some negative streaks in the matrix, colored blue in the figure. Such negative matrix elements $m_{ij} < 0$ are interference terms that reduce the contrast loss, meaning that the sum of intensities when pushing segments $i$ and $j$ individually gives a worse contrast than pushing them at the same time. This phenomenon is strongly correlated to the spatial frequency of the fringe pattern created by the segment pair. For a high spatial frequency (distant segments), the contrast degradation is averaged over the DH and is therefore minimized. For a low spatial frequency (close segments), the contrast can be degraded or improved due to the spatial configuration of the DH with respect to the fringe pattern.
We also show a simulated PASTIS matrix $M^{\text{sim}}$ in Fig.~\ref{fig:pastis-matrix_experimental}, right, calculated with the same calibration aberration per segment of $a_c = 40~nm$ WFE rms, but without any WFE or measurement noise in the optical system. We can see that the general morphology of the simulated and experimental matrices is the same -- in particular, it is the same segment pairs that show the highest and lowest contrast contribution in the image plane, relatively speaking. The experimental matrix is noisier though, and it has a slightly higher overall amplitude. Here, we want to show that it is feasible to directly measure an experimental PASTIS matrix, which will represent the real optical system more accurately, and compare this to results obtained with the simulated matrix.

\subsubsection{Contrast model validation}

\begin{figure}
    \resizebox{\hsize}{!}{\includegraphics{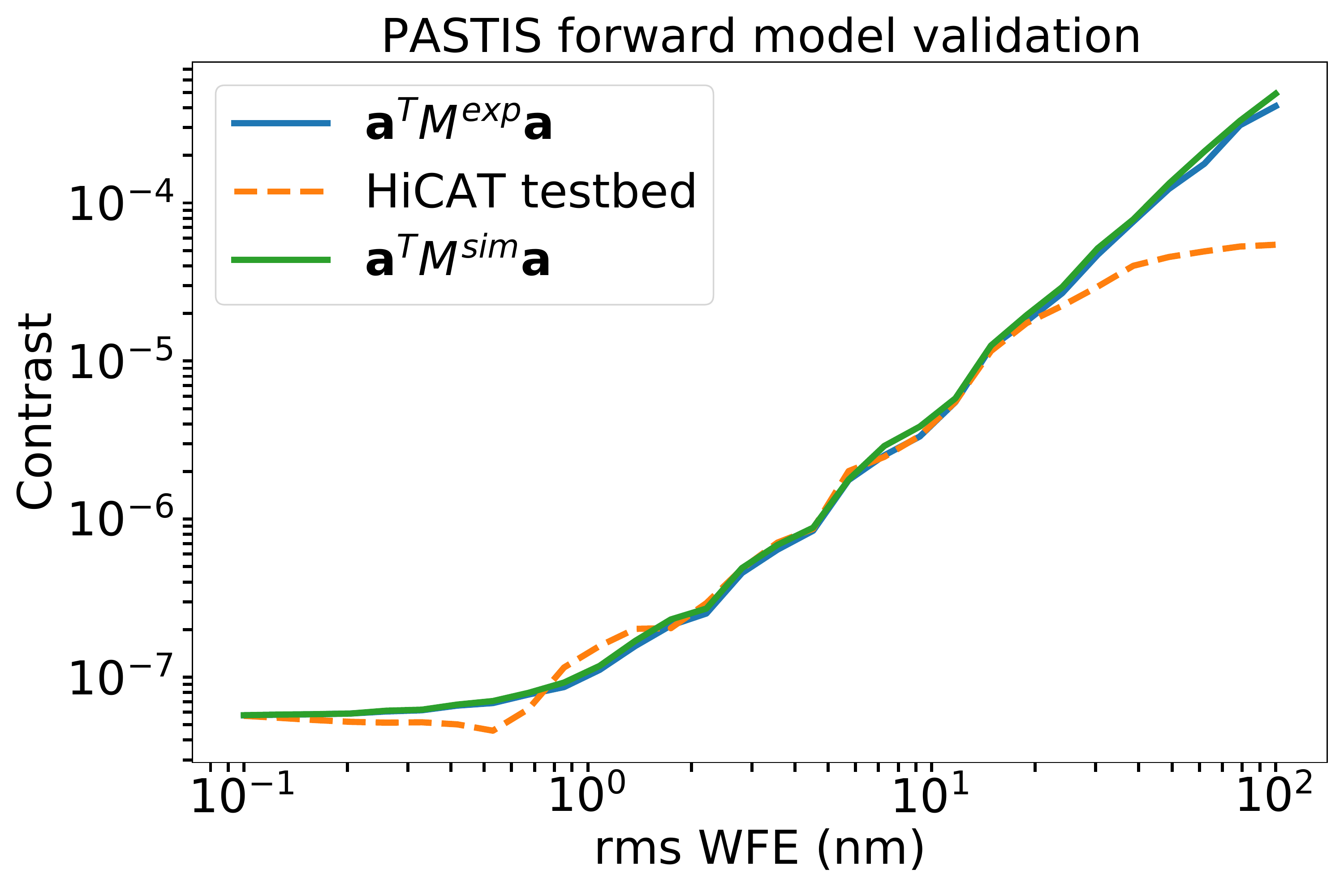}}
   \caption[Hockey stick validation] 
   {\label{fig:hockeystick_experimental} 
    Validation of the deterministic forward model in Eq.~(\ref{eq:pastis-equation_exp-paper}) by computing the contrast from the same segmented WFE maps with the experimental PASTIS matrix with Fig.~\ref{fig:pastis-matrix_experimental}, middle (solid blue), simulated PASTIS matrix from Fig.~\ref{fig:pastis-matrix_experimental}, right (solid green), as well as applying them on the HiCAT testbed (dashed orange). The curves flatten out to the left, at the coronagraph floor $c_0$, and show linear behavior at increasing WFE, giving them their hockey stick-like shape. The contrast calculation from the PASTIS equation with both matrices shows very good accordance with the testbed measurements, all three lines overlap at WFE larger than 1~nm rms.}
   \end{figure}

To validate the instantaneous PASTIS forward model, we compare the contrast for a segmented phase error calculated with Eq.~(\ref{eq:pastis-equation_exp-paper}), against the contrast measured for the same segmented phase map applied to the IrisAO on the testbed. For a range of rms WFE values, we generate a random segmented phase map $\mathbf{a}$, and we scale it to a given global rms WFE. Then we evaluate Eq.~(\ref{eq:pastis-equation_exp-paper}) both with the simulated matrix $M^{\text{sim}}$ and with the experimental matrix $M^{\text{exp}}$, and measure the resulting average DH contrast on the testbed. The results are plotted in Fig.~\ref{fig:hockeystick_experimental}. We observe that the results from the PASTIS equation using the experimental matrix (solid blue) show very good accordance with the testbed measurements (dashed orange), the curves overlap at a global rms WFE beyond 2~nm. The contrast calculated with the simulated matrix (solid green) yields an equally accurate result compared to the hardware measurements. We can clearly see all curves flatten out towards the left, where they are limited by the coronagraph floor, producing a hockey stick-like shape.

\subsection{Validation of mode contrast allocation}
\label{subsec:mode-tolerances_exp_paper}

\begin{figure*}
    \centering
    \includegraphics[width=17cm]{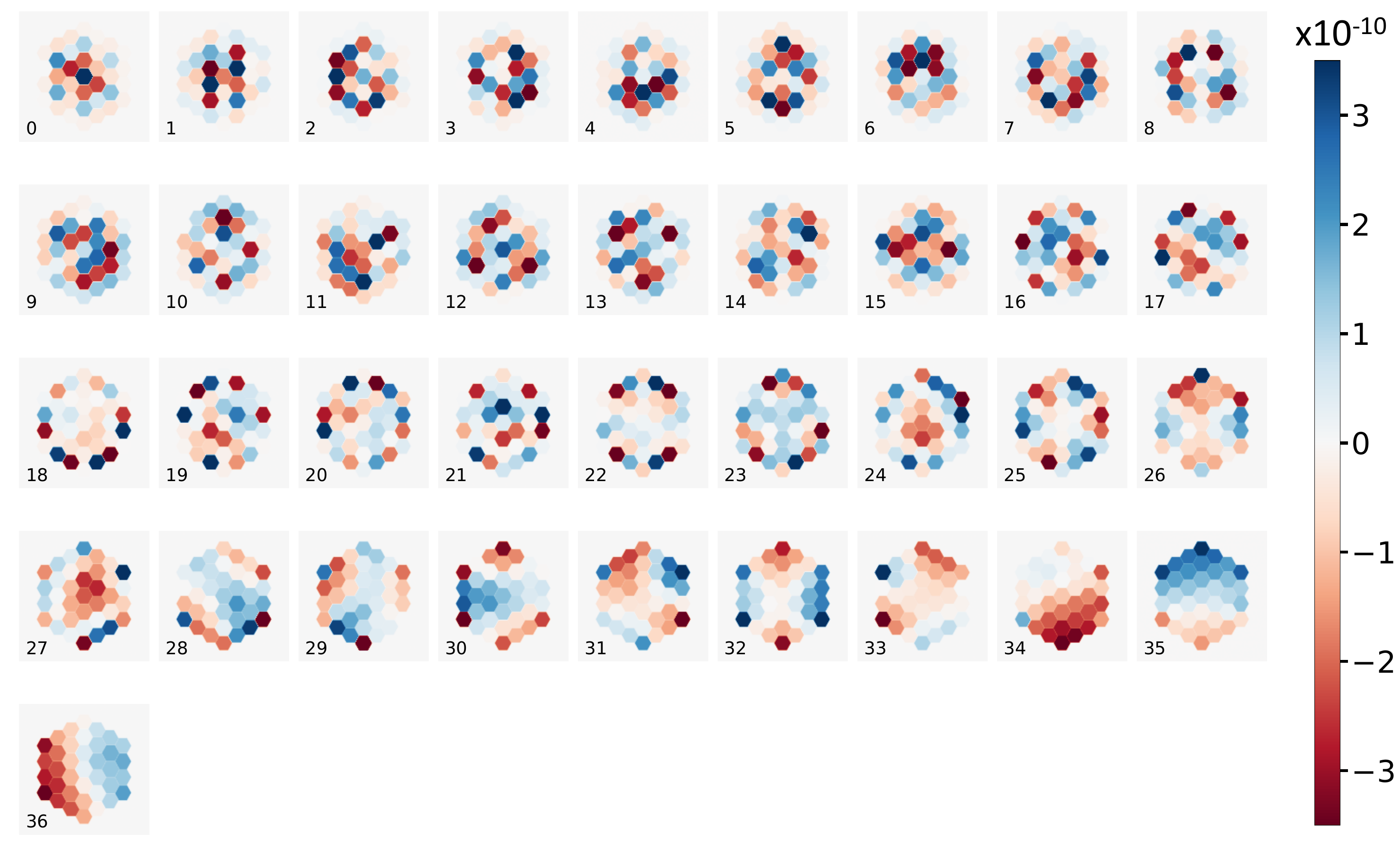}
    \caption{All experimental PASTIS modes for HiCAT with a CLC, for local piston aberrations, sorted from highest to lowest eigenvalue. The modes are unitless, showcasing the relative scaling of the segments to each other, and between all modes. They gain physical meaning when multiplied by a mode aberration amplitude $\sigma_{b_p}$ in units of wavefront error or phase. Their relative impact on final contrast is given by their eigenvalues.}
    \label{fig:postage-stamp-modes_experimental}
    \end{figure*}

We proceed with an eigendecomposition of the experimental PASTIS matrix \citep[Sect.~3]{Laginja2021AnalyticalTolerancingSegmented} and calculate its eigenmodes, shown as the optical PASTIS modes in Fig.~\ref{fig:postage-stamp-modes_experimental}. The modes are ordered from highest to lowest eigenvalue, indicating their comparative impact on the DH average contrast in their natural normalization.
These segmented PASTIS optical modes for the HiCAT testbed with a CLC represent the modal contrast sensitivity of the instrument with respect to segment misalignments, with the sensitivity quantified by their respective eigenvalues. As shown previously, the lowest-impact modes (bottom of Fig.~\ref{fig:postage-stamp-modes_experimental}) are dominated by low-spatial frequency components that are similar to discretized Zernike modes. The two modes with indices 35 and 36 in particular, the lowest-sensitivity modes, represent orthogonal tip and tilt modes over the entire pupil. High-impact modes (top of Fig.~\ref{fig:postage-stamp-modes_experimental}) display high-spatial frequency content, mostly in the central area of the pupil which is unconcealed by the LS.

\begin{figure}
    \resizebox{\hsize}{!}{\includegraphics{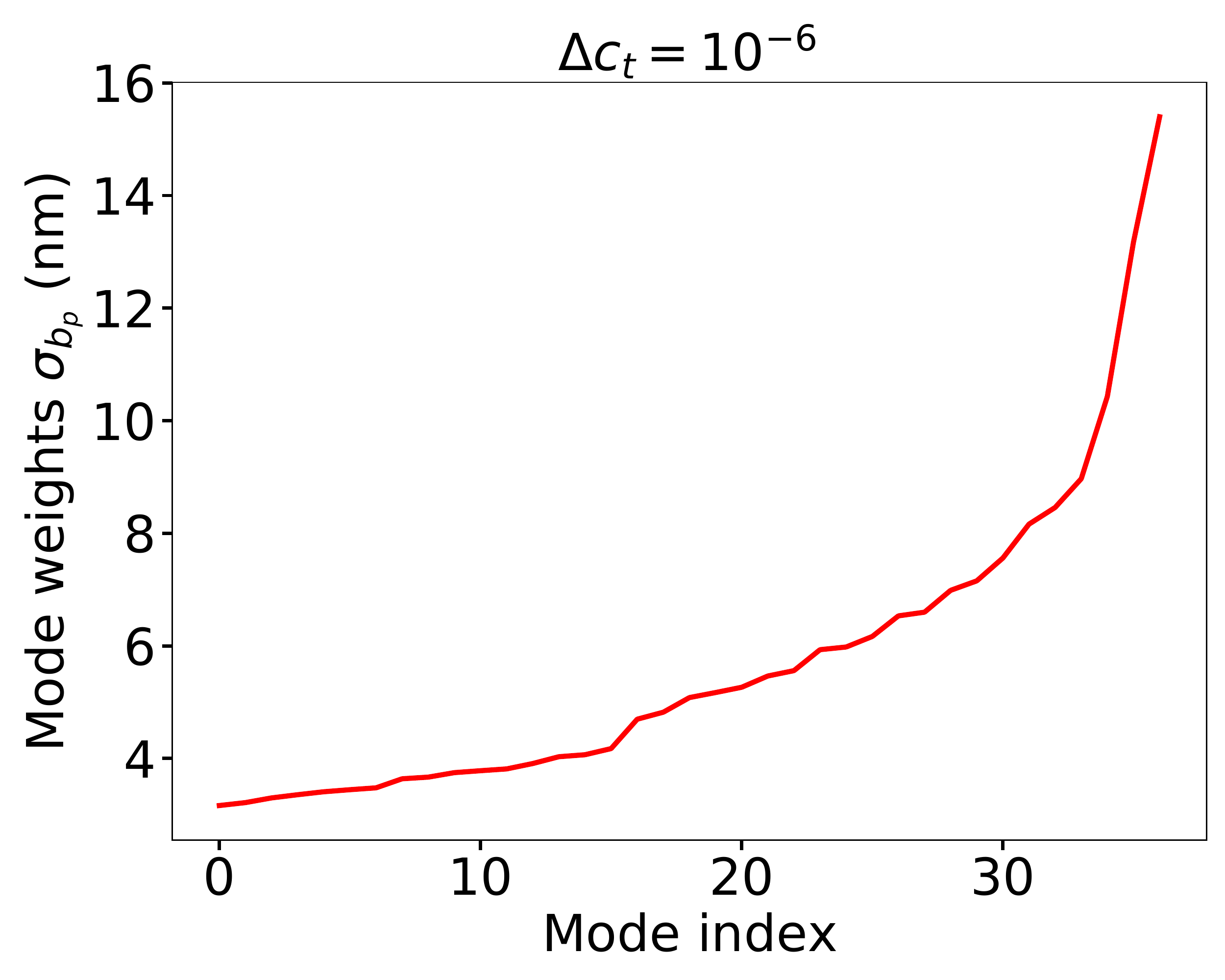}}
   \caption[Mode requirements] 
   {\label{fig:mode-requirements_experimental}
    Mode requirements as calculated with Eq.~(\ref{eq:mode-reqs-delta-ct_exp-paper}) for a uniform contrast contribution per mode, to a differential target contrast of $10^{-6}$. These are used as modal weights in the cumulative contrast measurement shown in Fig.~\ref{fig:cumulative-contrast-experimental}.}
   \end{figure}

\begin{figure}
    \resizebox{\hsize}{!}{\includegraphics{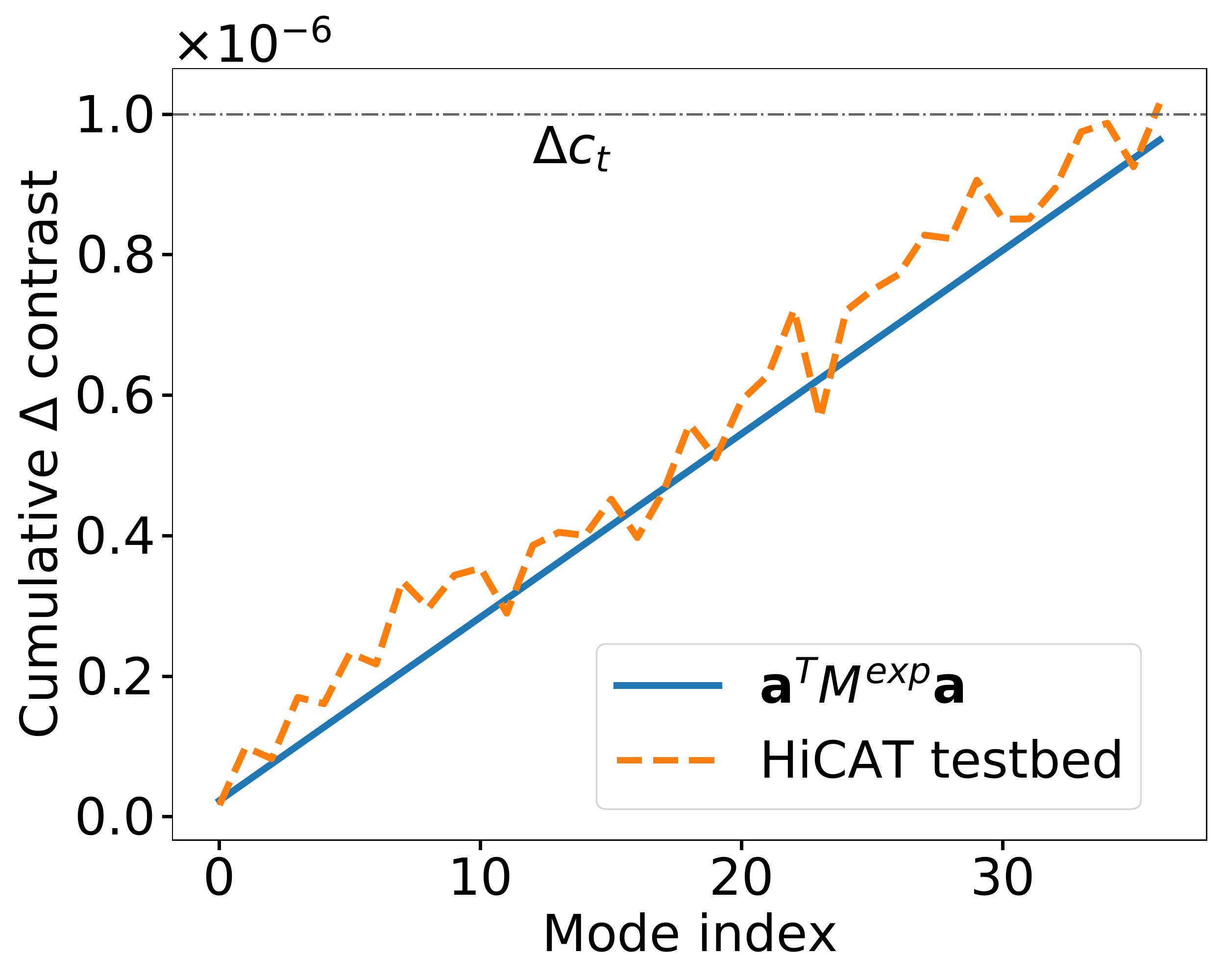}}
   \caption[Cumulative contrast plot] 
   {\label{fig:cumulative-contrast-experimental}
    Cumulative contrast plot for the uniform mode requirements shown in Fig.~\ref{fig:mode-requirements_experimental}, calculated both with the PASTIS forward model in Eq.~(\ref{eq:pastis-equation_exp-paper}), with the experimental PASTIS matrix (solid blue), and measured on the HiCAT testbed (dashed orange). The analytical PASTIS calculation shows the perfect linear curve that the uniformly scaled modes would cause on an ideal testbed without noise and drifts. The testbed measurements show noisy behavior around the ideal curve, which are likely caused by segmented DM command uncertainties for small actuations and an insufficiently well calibrated coronagraph floor.}
   \end{figure}

The PASTIS modes in Fig.~\ref{fig:postage-stamp-modes_experimental} form an orthonormal mode basis, making them independent from each other - each of them contributes to the overall contrast without influence from the other modes, see Eq.~(\ref{eq:delta-c-from-bps_exp-paper}). This can be used to define deterministic contrast allocations based purely on these optical modes \citep[Sect.~3.2]{Laginja2021AnalyticalTolerancingSegmented}. In the present example, we chose that each PASTIS mode should contribute uniformly to the total contrast, in which case we can calculate the exact mode weights for a particular target contrast:
    \begin{equation}
    \sigma_{b_p} = \sqrt{\frac{\langle \Delta c_t \rangle}{n_{modes} \cdot \lambda_p}}.
    \label{eq:mode-reqs-delta-ct_exp-paper}
    \end{equation}
In accordance to the formalism laid out in Sect.~\ref{sec:recall-pastis_exp_paper}, we make the mode tolerances independent of any given coronagraph floor by relating them to the differential target contrast $\Delta c_t$, displayed for a target contrast of $\Delta c_t = 10^{-6}$ in Fig.~\ref{fig:mode-requirements_experimental}. To validate the assumption of a contrast that is a simple sum of separate mode contributions, we run an experiment to measure the cumulative contrast of the deterministically scaled PASTIS modes. For this, we multiply the modes by their respective uniform requirement, $\sigma_{b_p}$, apply them cumulatively to the IrisAO, measure the resulting DH average contrast at each step and subtract the simultaneously measured coronagraph floor $c_0(t)$ from the results (Fig.~\ref{fig:cumulative-contrast-experimental}). We perform these propagations of the experimental eigenmodes both with the analytical PASTIS model in Eq.~(\ref{eq:pastis-equation_exp-paper}), using the experimentally measured matrix (solid blue), as well as with the HiCAT testbed (dashed orange).
In Fig.~\ref{fig:cumulative-contrast-experimental}, the cumulative measurements with HiCAT follow the general expected linear shape as displayed with the analytical PASTIS forward propagation using the experimental PASTIS matrix, but with some variations. These likely come from calibration errors in the segmented DM actuator influences for small displacements and contrast floor subtraction that is not accurate enough considering there will be no averaging effect during the comparatively short duration of this experiment ($\sim$5~min).

\subsection{Statistical validation of independent segment tolerances}
\label{subsec:segment-tolerances_exp_paper}

\subsubsection{Experimentally calibrated segment tolerances}

\begin{figure*}
   \centering
    \includegraphics[width = 17cm]{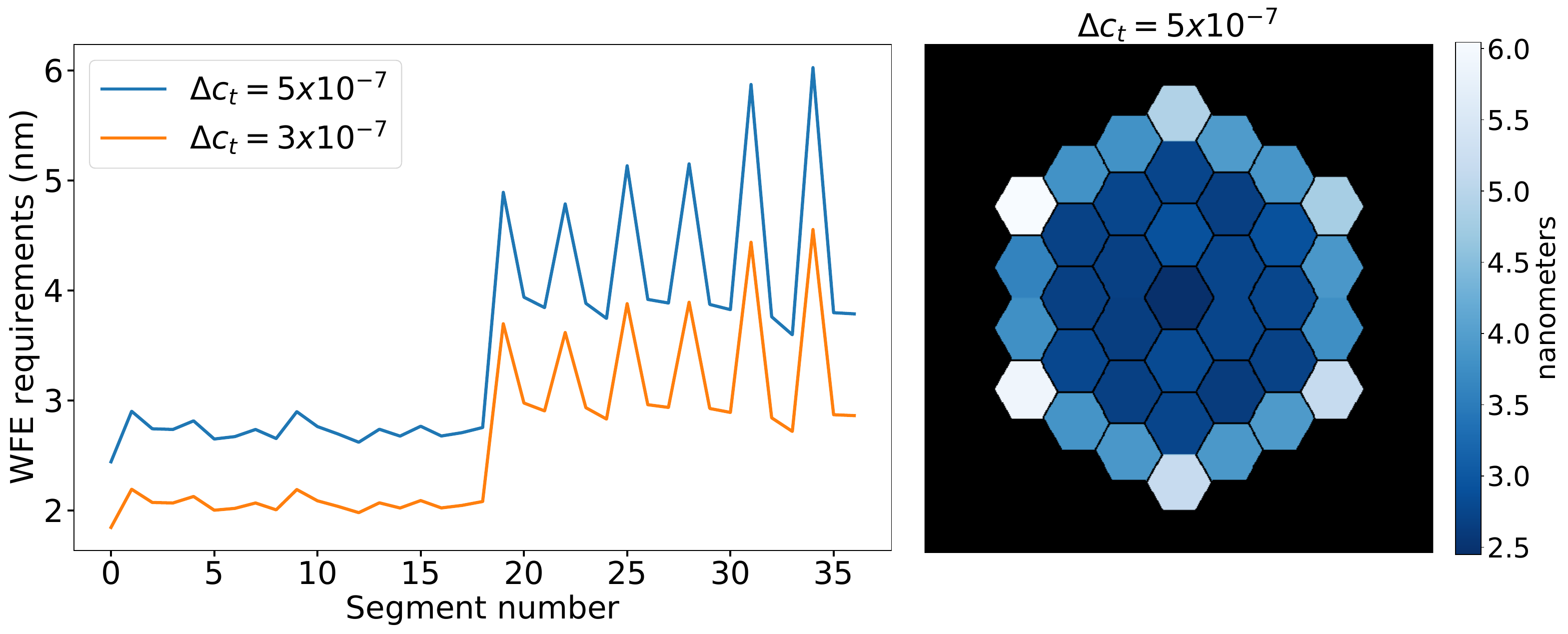}
   \caption[Segment requirements] 
   {\label{fig:segment-requirements_experimental} 
    \textit{Left:} Independent segment requirements as calculated with Eq.~(\ref{eq:mus-delta-c_exp-paper}) for target contrasts of $\Delta c_t = 5 \times 10^{-7}$ (top blue) and $\Delta c_t = 3 \times 10^{-7}$ (bottom orange). These numbers are the standard deviation of the tolerable WFE rms on each segment, if the target contrast is to be met as a statistical mean over many realizations of the segmented mirror. The range for the $5 \times 10^{-7}$ target spans from $2.5$ to $6~nm$. Both curves show a clear jump in the outermost ring (starting with segment number 19), which is highly concealed by the LS (see Fig.~\ref{fig:hicat_pupil_overlaps_experimental}). \textit{Right:} Tolerance map for a target contrast of $\Delta c_t = 5 \times 10^{-7}$, shown in blue on the left. The global left-right asymmetries are due to a slightly offset LS, which does not impact the contrast performance at this level, but is visible in the segment sensitivities. The map for $\Delta c = 3 \times 10^{-7}$ is a scaled version of the one shown here.}
   \end{figure*}

To fully validate the PASTIS tolerancing model for contrast stability, we calculate statistical segment-level requirements, from the experimental PASTIS matrix, for two target contrast values and measure their statistical contrast response with HiCAT. In this context, we use ``tolerances" and ``requirements" synonymously. In cases where the segments can be assumed to be independent from each other, as is the case for an IrisAO, we can calculate individual segment requirements \citep[Sect.~4.2]{Laginja2021AnalyticalTolerancingSegmented} with Eq.~(\ref{eq:mus-delta-c_exp-paper}) as a function of the differential target contrast. While the overall level of WFE requirements will be highly influenced by the Fourier filtering of the FPM, the different segments do not show uniform tolerance levels, as shown in Fig.~\ref{fig:segment-requirements_experimental}, left. These individual segment requirements are highly influenced by pupil features of the optical system. Looking at their spatial distribution in the HiCAT pupil, we can see in Fig.~\ref{fig:segment-requirements_experimental} (right) that the segments of the outer ring have more relaxed requirements than the two inner rings and the center segment. This is caused, in large part, by the LS, which is covering a large fraction of the segments in the outer ring because it is undersizing the pupil, which can be seen in Fig.~\ref{fig:hicat_pupil_overlaps_experimental}, left.
The two sets of segment-level WFE requirements displayed in Fig.~\ref{fig:segment-requirements_experimental}, left, represent a statistical description of the allowable WFE per segment if a delta target contrast of $5 \times 10^{-7}$ or $3 \times 10^{-7}$ is to be maintained as a statistical mean over many states of the segmented DM. We can observe how these two target contrast values yield segment requirements between 2.5 and 6~nm of WFE, which we know the IrisAO can reliably do; anything lower might lead to issues with the minimal stroke of the segmented DM and needs to be characterized in the future. Further, these particular delta contrast values are commensurable with the current testbed performance of HiCAT, staying just above the largest contrast fluctuations observed in Fig.~\ref{fig:contrast-drift-matrix-piston_experimental} ($\sim 1 \times 10^{-7}$), which are fast fluctuations that our model does not take into account.

\begin{figure*}
   \centering
   \includegraphics[width = 17cm]{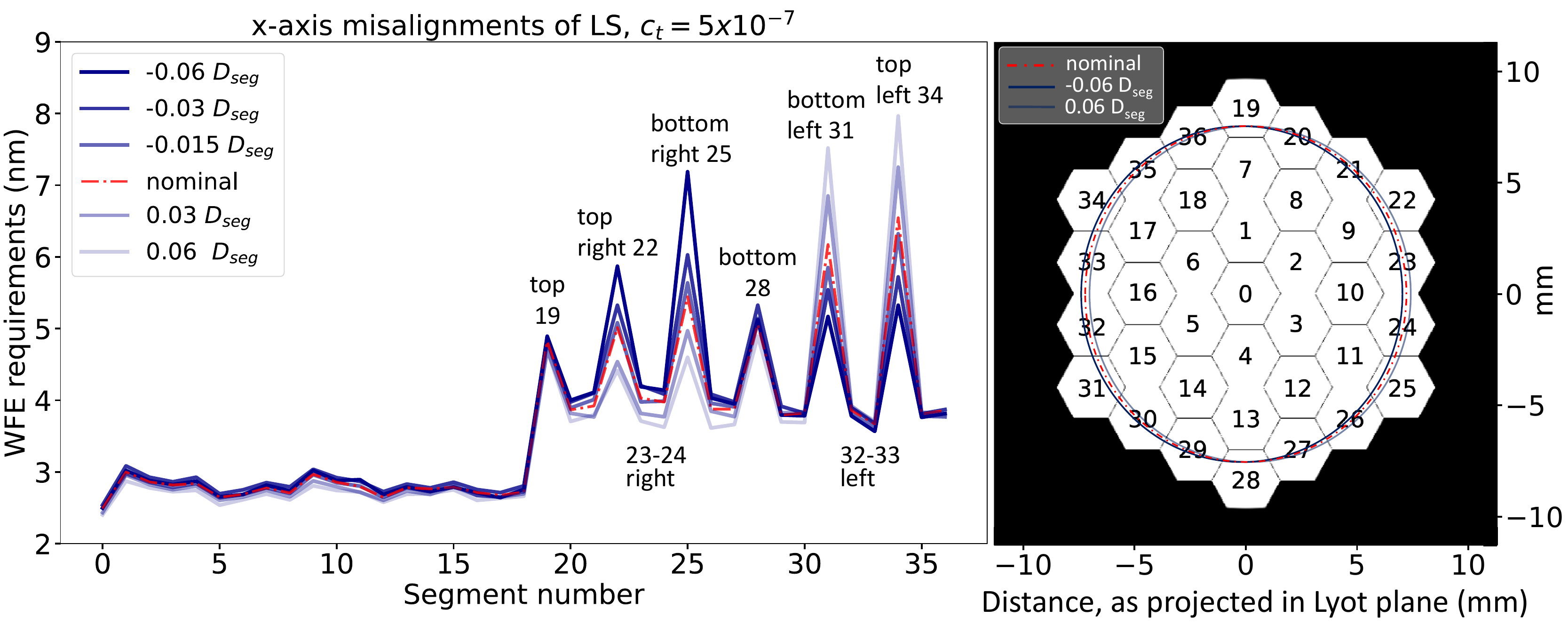}
   \caption[Sensitivity LS misalignment] 
   {\label{fig:sensitivity_LS_misalignment_experimental} 
    \textit{Left:} Individual segment requirements for a target contrast of $5 \times 10^{-7}$, measured for a range of x-axis misalignments of the LS. The legend is sorted in the same order like the lines for segment 25. The translation values are quoted in relation to the circumscribed segment diameter $D_{seg}=1.4$~mm, projected into the Lyot plane. The labels indicate select segments in the outer ring. \textit{Right:} Numbered segmented DM with overlaid LS positions. The nominal alignment is centered on the pupil (red dashed-dotted ellipse), with the dark and bright solid ellipses showing the maximum misalignments measured to either side. Details: see text.}
   \end{figure*}

As long as the components of the segment-level WFE on the DM follow independent zero-mean normal distributions whose standard deviations are given by the numbers in Fig.~\ref{fig:segment-requirements_experimental}, the target contrast will be recovered as the statistical mean over many such realizations. The tolerance map in Fig.~\ref{fig:segment-requirements_experimental}, right, shows a spatial representation of the segment-level standard deviations for $\Delta c = 5 \times 10^{-7}$; the map for $\Delta c = 3 \times 10^{-7}$ is a scaled version of the one plotted here and is not shown in this paper. The geometrical setup of the HiCAT pupil suggests a symmetry in the segment sensitivities along two axes, meaning for example that all four ``corner" segments should display the same tolerance level. The data in Fig.~\ref{fig:segment-requirements_experimental} underline this principal symmetry, but we do see a slight discrepancy between the corner segments on the left and right side. We attribute this to a slight left-right misalignment of the LS with respect to the IrisAO. To demonstrate this, we proceed by measuring the individual contrast sensitivities of all segments (only the PASTIS matrix diagonal, with Eq.~(\ref{eq:drifting-diagonal_exp-paper})) for varying lateral misalignments of the LS. Starting from the nominal centered alignment, we move the LS by a fraction of the segment size, characterized by its circumscribed diameter of $D_{seg}=1.4$~mm, as measured in the plane of the IrisAO. Then we run a couple of iterations of the WFS\&C algorithm in order to optimize the DH solution and recover the nominal contrast level, measure the average contrast response in the DH to imposed (individual) segment pokes of 40~nm of WFE, and use Eq.~(\ref{eq:mus-delta-c_exp-paper}) to calculate the WFE requirements per segment, for a target contrast of $\Delta c_t = 5 \times 10^{-7}$. Repeating this for five different misalignments along the x-axis yields the data shown in Fig.~\ref{fig:sensitivity_LS_misalignment_experimental}, left.
The nominal alignment case (dashed red) has been measured on a new DM solution for a DH contrast of $c_0 = 3.8 \times 10^{-8}$ (as opposed to the previous $c_0 = 2.5 \times 10^{-8}$), which is why the segment tolerances deviate slightly from the numbers indicated by the blue line in Fig.~\ref{fig:segment-requirements_experimental}. While the tolerances for segments on top and bottom of the pupil (19 and 28) show no changes across the different misalignments, all other segments of the outer ring do. These results are directly connected to the exposed area of a segment in each alignment state of the LS: the more area is exposed the more stringent its requirement turns out. This is most visible for the four corner segments (22, 25, 31 and 34), as their area doubles between the farthest alignment states. The values for the side segments (23, 24 on the right, and 32, 33 on the left) change as a function of misalignment too, but less so as their exposed area changes much less. The sensitivity analysis of single segments with respect to the position of the LS demonstrates the interest of the PASTIS tolerancing model: while the total WFE requirement, over the full pupil, is the same for all six LS states shown in Fig.~\ref{fig:sensitivity_LS_misalignment_experimental}, the segments display different individual tolerancing levels depending on where in the pupil they are located.

\begin{figure*}
   \centering
   \includegraphics[width = 17cm]{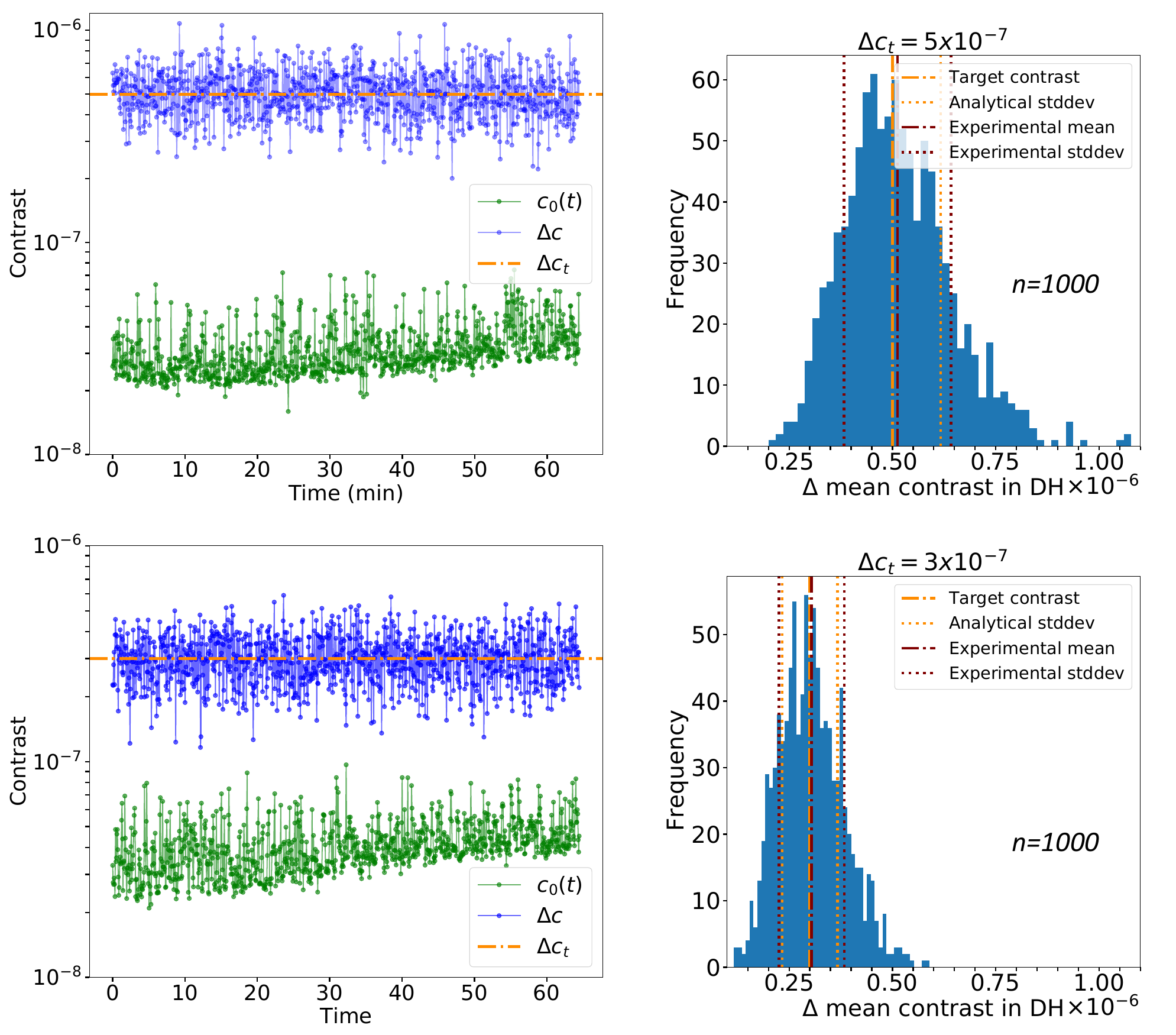}
   \caption[Monte Carlo segments] 
   {\label{fig:monte-carlo-segments_experimental} 
    Monte Carlo experiments on HiCAT to validate the independent segment error budget shown in Fig.~\ref{fig:segment-requirements_experimental}, calculated from the experimental PASTIS matrix with Eq.~(\ref{eq:mus-delta-c_exp-paper}), for target contrasts of $\Delta c_t = 5 \times 10^{-7}$ (top) and $\Delta c_t = 3 \times 10^{-7}$ (bottom). 
    \textit{Left:} Time series of the measured random contrasts (blue), from which the intermittently measured unaberrated contrast drift (green) has been subtracted. The total duration of each experiment is 65 min and captures 1000 randomly generated WFE maps. The target contrast of both experiments is indicated by the orange dashed lines.
    \textit{Right:} The contrast measurement from the random WFE to the left, plotted as histograms. The experimental distributions have a mean (red dashed-dotted lines) of $5.12 \times 10^{-7}$ (top) and $3.04 \times 10^{-7}$ (bottom). The expected standard deviations (orange dotted lines) as calculated by Eq.~(\ref{eq:var-of-c_exp-paper}) are $1.2 \times 10^{-7}$ (top) and $6.7 \times 10^{-8}$ (bottom), versus the experimentally measured standard deviations (red dotted lines) of $1.3 \times 10^{-7}$ and $7.7 \times 10^{-8}$.}
   \end{figure*}

In order to validate the computed per-segment tolerances, we proceed by running a Monte Carlo experiment for both target cases shown in Fig.~\ref{fig:segment-requirements_experimental}. For each experiment, we sequentially apply 1000 different WFE aberration patterns on the segmented DM and record the resulting average DH contrast values. The tolerances in Fig.~\ref{fig:segment-requirements_experimental} are the prescription determining how to draw these random WFE realizations: each segment-level WFE on segment $k$, in a single random WFE map $\mathbf{a}$, is drawn from its own zero-mean normal distribution with a standard deviation of $\mu_k$, given by Eq.~(\ref{eq:mus-delta-c_exp-paper}). This means that one random HiCAT WFE map is composed of 37 independent normal distributions with each a mean of zero, and a standard deviation of $\mu_k$, which then gets applied to the IrisAO on HiCAT, and a DH contrast measurement is made. Since the tolerancing target is $\Delta c_t$, which is independent of the coronagraph floor $c_0$, we intersperse measurements with a flat IrisAO in each iteration to capture the evolution of the contrast floor on the testbed. We subtract these $c_0(t)$ values from the DH measurements in the respective iteration in order to receive our final results in Fig.~\ref{fig:monte-carlo-segments_experimental}.
We plot the time series of the experimentally measured contrast responses from the segmented WFE maps in the two left panels in Fig.~\ref{fig:monte-carlo-segments_experimental}. The green bottom curves depict the evolution of the contrast floor over time, with the IrisAO DM repeatedly being reset to its best flat. The top blue curves show the contrast measurements with random WFE map realizations applied to the segmented DM, as prescribed with the tolerances in Fig.~\ref{fig:segment-requirements_experimental}, from which the green contrast floor has been subtracted. The histograms of the blue contrast curves are plotted in the two right panels in Fig.~\ref{fig:monte-carlo-segments_experimental}.

The resulting histograms have experimentally measured mean values of $5.12 \times 10^{-7}$ and $3.04 \times 10^{-7}$ (red dashed-dotted lines). These are very close to the target contrasts of $5 \times 10^{-7}$ and $3 \times 10^{-7}$ (orange dashed-dotted lines). This excellent fit between the target and experimental values ($\sim$1--3\%) is clearly sufficient to prove our concept at the required level. The difference is larger for the standard deviations: the experimentally measured values of $1.3 \times 10^{-7}$ and $7.9 \times 10^{-8}$ (red dotted lines) are off from their analytically calculated values with Eq.~(\ref{eq:var-of-c_exp-paper}) of $1.2 \times 10^{-7}$ and $6.7 \times 10^{-7}$ (orange dotted lines), by 5--7\%. This is expected because of a slower convergence of variance estimators with respect to mean estimators. We note a slight asymmetry in the histograms, biased towards higher contrast. While the underlying assumption for the tolerancing method used in this paper is that the segment aberrations follow Gaussian statistics \citep[Sect.~4.1]{Laginja2021AnalyticalTolerancingSegmented}, which is reasonable for the aberrations on a segmented mirror telescope, the contrast itself does not follow a Gaussian statistic; it is the sum of squared Gaussian variables which is also known as a generalized $\chi^2$ statistic, causing the asymmetry.

\subsubsection{Simulated segment tolerances}

\begin{figure}
    \resizebox{\hsize}{!}{\includegraphics{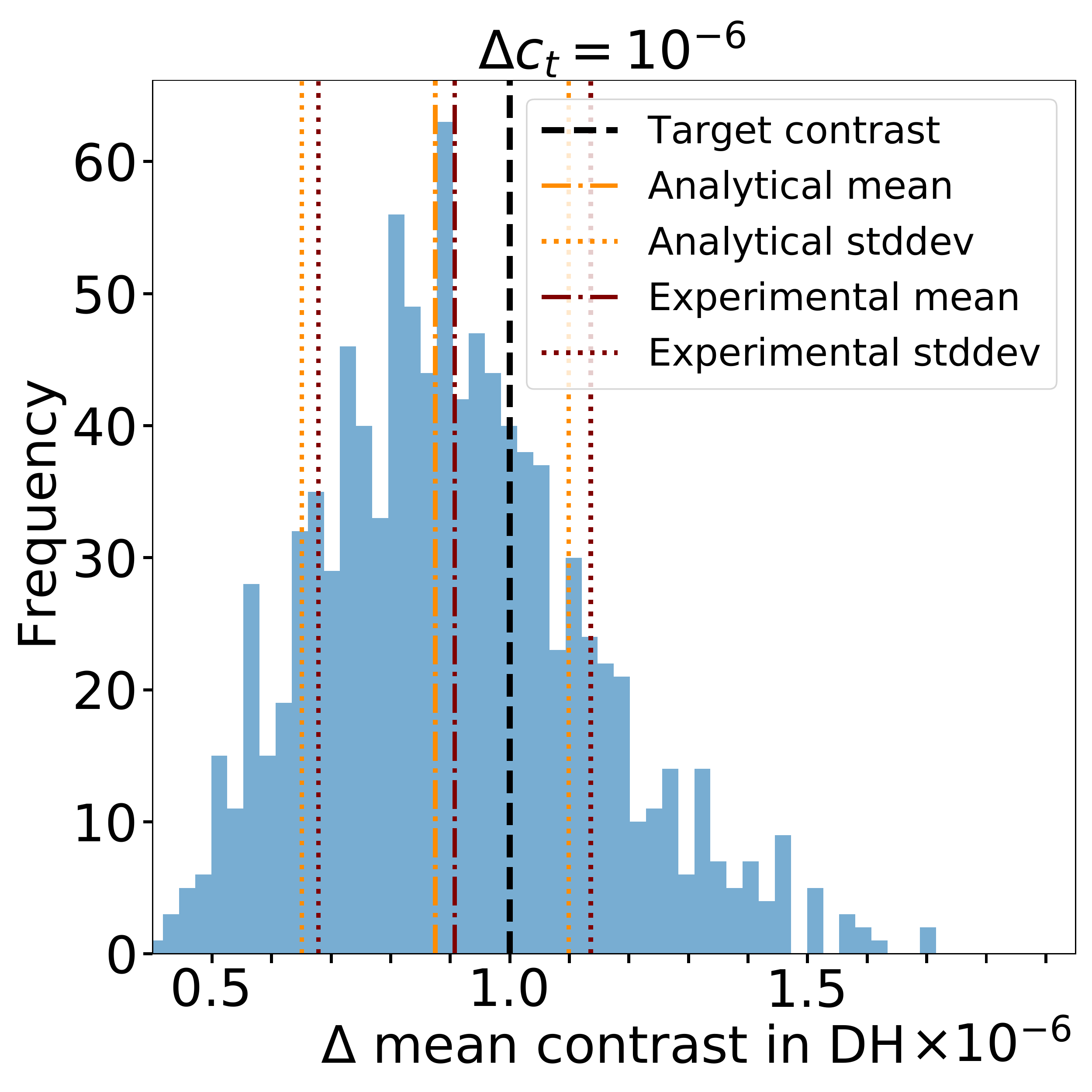}}
   \caption[Monte Carlo segments from simulated tolerances] 
   {\label{fig:monte-carlo-segments-sim-tolerances_experimental} 
    Monte Carlo experiment on HiCAT for segment tolerances obtained from the simulated PASTIS matrix diagonal using Eq.~(\ref{eq:mus-delta-c_exp-paper}). The target contrast $\Delta c_t = 10^{-6}$ as determined by the simulated tolerances is indicated with the black dashed line. The experimental mean over 1000 segmented WFE maps (red dashed-dotted line) does not recover the target contrast. But we are able to predict this discrepancy analytically: using Eqs.~(\ref{eq:mean-delta-c_exp-paper}) and (\ref{eq:var-of-c_exp-paper}), we can calculate the resulting contrast mean (orange dashed-dotted line) and variance (orange dotted lines) from the experimental PASTIS matrix and simulated segment tolerances directly. Details: see text.}
   \end{figure}

Ideally, we could use a simulated model directly to be able to assess the segmented tolerancing levels for a particular contrast level. To this end, we use $M^{\text{sim}}$ to calculate a set of segment tolerances for a target contrast of $10^{-6}$ and repeat the experiment described above, with 1000 iterations. Most efforts put into the simulator to date were intentionally invested in matching the operational interface, and the optical scales and morphology (e.g., orientations, sampling and location of diffraction features, photometry). The optical model currently matches its contrast predictions to results on the hardware to within a factor of a few, which is why in this experiment, we chose a less demanding target contrast compared to the experiments shown in Fig.~\ref{fig:monte-carlo-segments_experimental}. In this way, we retain an error margin for the resulting hardware contrast values that takes this into account. The results of this experiment are plotted in Fig.~\ref{fig:monte-carlo-segments-sim-tolerances_experimental}.
We can see how the experimentally measured mean contrast (red dashed-dotted line) fails to meet the target contrast marked by the black dashed line with an error of 10\%. This indicates that the tolerances derived from the simulated PASTIS matrix are less accurate than the experimentally measured ones.

\begin{figure}
    \resizebox{\hsize}{!}{\includegraphics{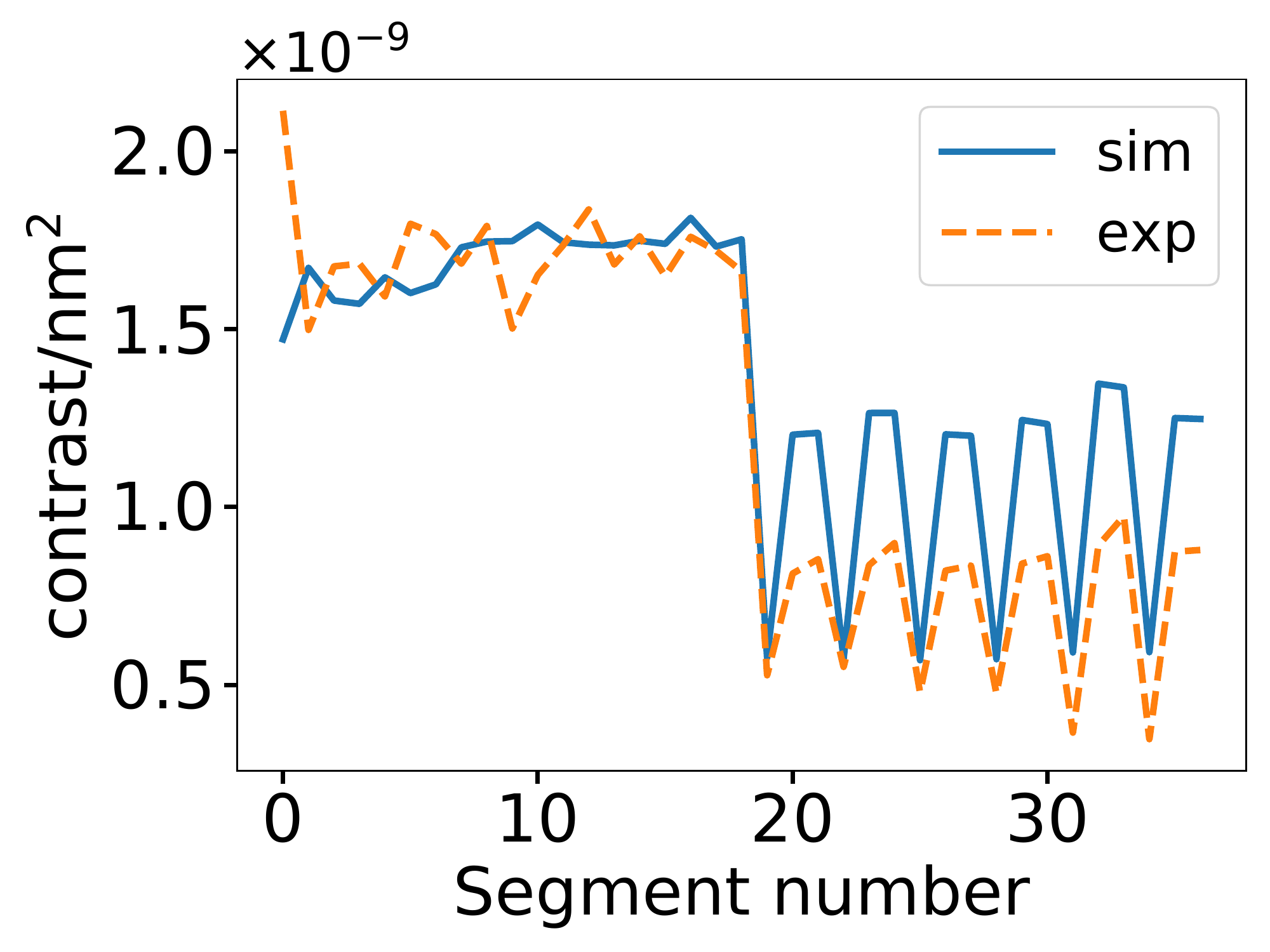}}
   \caption[PASTIS matrix diagonals exp and sim] 
   {\label{fig:matrix-diagonals_experimental} 
    Comparing the PASTIS matrix diagonals from the hardware matrix $M^{\text{exp}}$ (orange dashed line) and simulated matrix $M^{\text{sim}}$ (blue solid line). Segment number 0 (center segment) in the experimental matrix has the largest absolute contrast influence, but the simulated matrix assumes on average the larger influence of the individual segments on the average DH contrast. This translates directly into more stringent segment tolerances $\mu_k$ than necessary for the given hardware setup.}
   \end{figure}

Since the independent segment requirements in Eq.~(\ref{eq:mus-delta-c_exp-paper}) (case of diagonal covariance matrix $C_a$) only depend on the PASTIS matrix diagonal, we compare the diagonals of the experimentally measured and simulated PASTIS matrix in Fig.~\ref{fig:matrix-diagonals_experimental}.
While the experimental matrix displays the higher absolute peak diagonal element (for the center segment), on average it contains lower contrast contributions per segment than the simulated matrix. In particular, the outer segments in the pupil are consistently over-estimated in terms of contrast contribution by the simulated matrix. A closer inspection of the pupil image shows that the segmented mirror model in the HiCAT simulator is in fact over-stretched along the x-axis compared to the real IrisAO on the testbed, by 1.4\%. As a consequence, the segments of the outer ring expose $\sim5$\% more total area compared to their visible area on the hardware, which in turn increases their influence on overall contrast in the simulator. With an increased contrast sensitivity to these segments in simulation, it explains why the Monte Carlo experiment in Fig.~\ref{fig:monte-carlo-segments-sim-tolerances_experimental} misses the targeted mean contrast: since the simulated matrix assumes a larger contrast influence by each individual (outer ring) segment, and the segment tolerances $\mu_k$ are inversely proportional to the matrix, the resulting per-segment requirements turn out more restrictive than they need to be in reality. This result shows that the simulated and experimentally measured PASTIS matrices have significant differences on their respective diagonals, which is not captured in Fig.~\ref{fig:hockeystick_experimental}. This is because the statistical analysis in Fig.~\ref{fig:monte-carlo-segments-sim-tolerances_experimental} uses only the matrix diagonal, while the deterministic model in Fig.~\ref{fig:hockeystick_experimental} (Eq.~(\ref{eq:pastis-equation_exp-paper})) uses the full PASTIS matrix. In the latter case, it turns out that the differences between $M^{\text{exp}}$ and $M^{\text{sim}}$ average out.

While the discrepancies between the experimental and simulated matrix lead to an offset in the tolerancing results that can be improved upon with a more accurate model, we can show that this offset is directly predictable by using the experimentally measured matrix $M^{\text{exp}}$ with Eqs.~(\ref{eq:mean-delta-c_exp-paper}) and (\ref{eq:var-of-c_exp-paper}). We consider the segment tolerances obtained with $M^{\text{sim}}$ to be a covariance matrix $C_a^{sim}$, with the segment variances filling the covariance matrix diagonal. This allows us to calculate the analytically predicted mean with $\tr(M^{\text{exp}} C_a^{sim}) = 8.75 \times 10^{-7}$ (orange dashed-dotted line), and the variance with $2 \tr [(M^{\text{exp}} C_a^{sim})^2]$, with its square root yielding a standard deviation on the contrast of $ 2.2 \times 10^{-7}$ (orange dotted lines). These values accord with the experimentally measured mean of $ 9.07 \times 10^{-7}$ (red dashed-dotted line) and standard deviation of $ 2.3 \times 10^{-7}$ (red dotted lines) to within a statistical error. Usage of these formulae circumvents the need to re-evaluate the contrast response of a large number of segmented aberration maps when a new segment covariance matrix $C_a$ is obtained from modeling, and can be used for the direct assessment of the contrast performance for by-design instruments.

Overall, our experiments on HiCAT present successful experimental validations of the PASTIS model for a specific high contrast instrument. We measured an experimental PASTIS matrix and validated it by comparing its modeled contrast results to testbed measurements. We decomposed the matrix into independent optical modes that we scaled uniformly and cumulatively to a differential target contrast of $10^{-6}$. We calculated statistical segment-level WFE tolerances under the assumption of independent segments and validated them with Monte Carlo experiments at target contrasts of $5 \times 10^{-7}$ and $3 \times 10^{-7}$, measuring the contrast from randomly drawn segmented WFE maps as prescribed by the derived requirements. While using tolerances derived from a simulated PASTIS matrix was not enough to reach a particular contrast goal, the experimentally measured matrix allowed us to validate the analytical contrast predictions in terms of a contrast mean and variance for an arbitrary segment covariance matrix. Future work will aim to optimize the differing segment sizes in the segmented DM model; however, matching the contrast prediction of the HiCAT simulator to the hardware is out of the scope of this study.


\section{Discussion}
\label{sec:discussion_exp_paper}

The astronomical community's experience with space-based segmented observatories is currently limited to JWST, which will be launched very soon. Because of gravity and thermal constraints, the telescope was not aligned entirely on the ground to test its optical performance. Instead, a large number of ground optical tests including interferometry and other metrology techniques were performed to validate the observatory-level optical model, including for example radius of curvature, or interferometric alignment of adjacent segments \citep{Perrin2018UpdatedOpticalModeling,Knight2012ObservatoryAlignmentJames}. Something like LUVOIR would be two to three times larger than JWST and therefore it is anticipated that performance validations of the observatory will rely heavily on model-based assessments due to the sheer size and complexity of the telescope.

This raises the necessity for modeling, performance prediction and tolerancing tools that are not dependent on testing of the fully integrated observatory system. The PASTIS tolerancing tool, presented in theory so far \citep{Laginja2021AnalyticalTolerancingSegmented,Leboulleux2018PairbasedAnalyticalModel}, provides us with methods to derive segment stability tolerances that in turn will determine the coronagraphic performance of an imaging instrument like those on LUVOIR. In this paper, we perform the first hardware validation of an experimentally calibrated tolerancing model. We use this to determine the constraints on segmented mirror stability for various levels of contrast on the HiCAT testbed, which can be used to derive requirements of future testbeds that will perform system analyses for the LUVOIR mission. This paper validates the statistics of phasing errors without making any time scale assumptions for the WFE. Therefore, this applies not only to static phasing residuals (which can inform the design of a sensing and control strategy), but also to dynamic tolerances in the context of adaptive optics (AO) loops.

During the experiments, we observed that the variability in the static contrast poses a problem to the accurate measurement of the experimental PASTIS matrix for the mirror segments, and the tolerancing validations. We solved this issue in good part by adopting the differential contrast $\Delta c$ as the metric of interest: We have introduced a reformulation of the PASTIS tolerancing formalism in order to separate the contrast influence of segmented mirror misalignments from all other aberration sources in the instrument (see Sect.~\ref{sec:recall-pastis_exp_paper}). This is particularly useful when we assume a time-dependent phase aberration term $\phi_{ab}(\mathbf{r}, t)$ which will influence the absolute raw contrast.

We successfully validated the PASTIS model on a real high contrast instrument by measuring an experimental PASTIS matrix $M^{\text{exp}}$. We performed the individual segment tolerancing with this matrix and have shown that the derived segment requirements are indeed the correct standard deviations for a targeted mean contrast. The experimental results in Fig.~\ref{fig:monte-carlo-segments_experimental} coincide very well with the analytical formulas for contrast mean and variance in Eqs.~(\ref{eq:mean-delta-c_exp-paper}) and (\ref{eq:var-of-c_exp-paper}), down to a statistical error.

\begin{figure}
    \resizebox{\hsize}{!}{\includegraphics{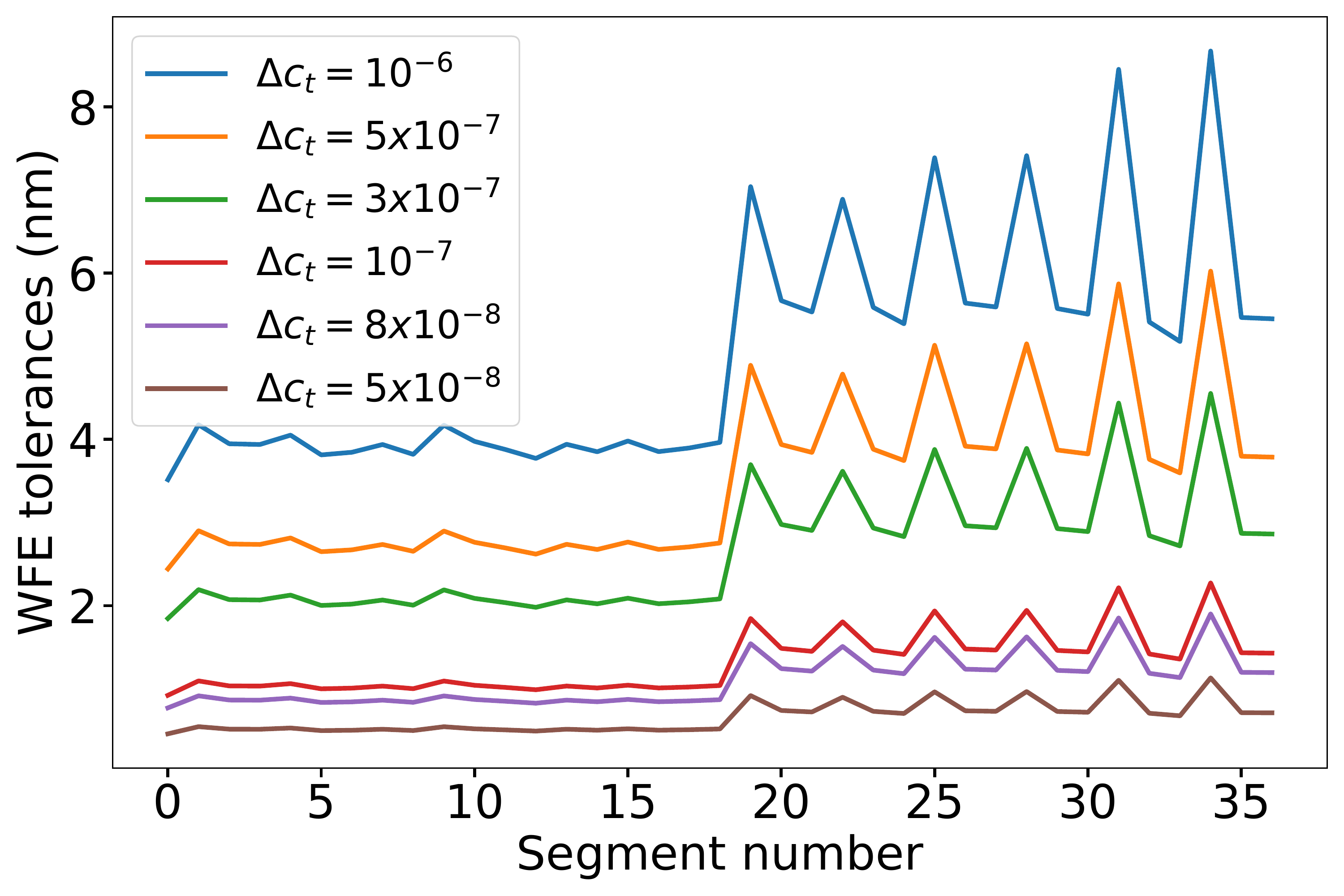}}
    \caption{Comparing independent segment WFE tolerances for different levels of differential target contrast, calculated from the experimental PASTIS matrix $M^{\text{exp}}$. The tolerances for $\Delta c_t = 5 \times 10^{-7}$ and $\Delta c_t = 3 \times 10^{-7}$ have been validated experimentally in the current performance regime on HiCAT. The other lines describe predicted requirement limits on the segmented mirror for deeper contrast levels. These curves isolate the influence of segmented mirror drifts from other system components that influence the contrast. The curves appear top to bottom in the same sequence like in the figure legend.}
    \label{fig:compare-mus_experimental}
    \end{figure}

We can use this method to extrapolate the stability requirement of the segmented mirror to contrast levels that lie beyond the current performance capability of the testbed. In the case of HiCAT, we are currently limited to a static contrast of $2 \times 10^{-8}$, and contrast variations up to the order of $1 \times 10^{-7}$ due to influences from the testbed environment and an incoherent background attributed to the light source. After validating the individual segment requirements in a contrast regime in which the contrast floor and drift do not pose any problems, we compare the individual segment tolerances as calculated with $M^{\text{exp}}$ in Eq.~(\ref{eq:mus-delta-c_exp-paper}) for various target contrasts down to the $10^{-8}$ regime in Fig.~\ref{fig:compare-mus_experimental}.
This result informs us about the level of the segmented mirror contrast stability for future demonstrations that intend to get closer to the envisioned LUVOIR performance. For example, the segmented mirror needs to be able to keep a sub-nanometer stability if we want to achieve a contrast of $10^{-8}$. The numbers in Fig.~\ref{fig:compare-mus_experimental} exclusively show the requirements on the WFE component caused by the segmented mirror, which builds a fundamental piece in the total error budget of a segmented high contrast instrument.

The mean differential contrast $\langle\Delta c\rangle$ can be written as a sum of unrelated contributions from the segments only if the segment pistons are statistically independent (see Eq.~(\ref{eq:get-muk-from-matrices_exp-paper})), that is, if the matrix $C_a$ is diagonal. But even if the segment pistons are correlated and even if these correlations are unknown, the mean differential contrast \textit{can} be written as a sum of independent contributions, namely those of the optical modes of the system, see Eq.~(\ref{eq:delta-c-from-bps_exp-paper}).

We showed that it is possible to use Eq.~(\ref{eq:mode-reqs-delta-ct_exp-paper}) in order to scale the individual optical modes to yield a particular target contrast (Fig.~\ref{fig:cumulative-contrast-experimental}). While there remain deviations from the exact equal contrast contribution of each mode to the overall contrast, this result clearly demonstrates that we can impose a weighted scaling on these modes in order to influence the contrast. The remaining errors we see in the uniform mode tolerancing are likely coming from calibration errors in the segmented DM actuator influences, unidentified aberrations in the system, and an insufficiently calibrated contrast floor for the short duration of this experiment ($\sim$5~min).

Fundamentally, these optical PASTIS modes represent the natural instrument modes with respect to contrast, which can be exploited for closed-loop adaptive optics (AO control during the operation of the instrument). The LUVOIR mission aims to deploy a full AO system in space, which requires control on all spatial frequencies. While fast, low-spatial frequency control can be done with a low-order wavefront sensor (LOWFS), a high-order wavefront sensor (HOWFS), which is not subjected to spatial filtering, will be sensing aberrations on the segmented pupil \citep{Pueyo2021spietalk,Pogorelyuk2021InformationtheoreticalLimitsRecursive}. 

Optimal wavefront control strategies will need to optimize the DH contrast, which is the main science metric for these future instruments.  The PASTIS modes therefore offer a natural application to this problem since their contrast influence is directly quantified by their respective eigenvalues. PASTIS modes should therefore be investigated further as part of the design of the High Order modal control scheme of a multiple-layer space AO system for high contrast applications. 

There are existing methods \citep{Chambouleyron2021VariationZernikeWavefront} that allow to design a Fourier-filtering WFS with given sensitivity to specific modes. Exploiting this, one could design a WFS focal-plane mask (thus a transmissive mask) giving an enhanced sensitivity to the modes degrading the contrast (whose influence mainly falls inside the DH area), accepting to have a reduced sensitivity outside. With such a WFS, the control loop could be optimized to maximize the contrast performance of the instrument directly.

\section{Conclusions}
\label{sec:CONCLUSIONS_exp_paper}

Accurate tolerancing of different WFE contributions on future large observatories is crucial in order to be able to design systems capable of stable enough contrast levels for exoEarth detection, to predict their contrast stability and assess their performance. In particular, the WFE contributions from cophasing errors on segmented telescopes will have a direct impact on the performance of the high contrast instrument. In this paper, we used the PASTIS tolerancing model to perform a segmented WFE tolerancing analysis on the HiCAT testbed, and presented experimental validations to demonstrate its utility.

We successfully measured an experimental PASTIS matrix on a 37-segment IrisAO mirror after isolating the influence of the segments in the overall contribution to contrast drift. The individual segment tolerances calculated from this matrix yield an accurate mean contrast and variance in Monte Carlo experiments when compared to analytical predictions, up to a minimal statistical error. We also compared these experimentally obtained segment tolerances to equivalent results obtained from a simulated PASTIS matrix. The experimental measurements were more accurate for performance predictions, but the errors from the simulated segment tolerances can likely be minimized with a more accurate model of the segmented mirror.

Combining the experimental PASTIS matrix, representing the realistic contrast influence of the testbed, with a covariance matrix describing segment piston variations allowed us to correctly predict the resulting contrast mean and variance. This allows for a simple evaluation of the expected contrast stability of a per-design instrument. Such a covariance matrix can be a diagonal one to describe the independent segments from a simple segmented mirror design, as in the HiCAT case, or a non-diagonal one that incorporates knowledge from opto-mechanical correlations between segments coming from realistic finite-element modeling.

We use the experimentally measured matrix to predict the required WF stability of the segmented DM on HiCAT for contrast levels that are currently out of reach due to environmental influences on the testbed. We first validate the segment tolerances for a differential contrast of $5 \times 10^{-7}$ and $3 \times 10^{-7}$, for which the segment requirement standard deviations range from 2.5 to 6~nm. We then proceed by establishing a set of requirements for the segmented mirror for a contrast contribution in the $10^{-8}$ regime, and conclude that the WF stability from the segmented DM will have to be better than 1~nm for each individual segment.

Our future work aims to measure a simulated PASTIS matrix with a more accurate model and use it to derive WFE tolerances that correctly define the stability requirements for a target mean contrast measured on the hardware. Further, we aim to demonstrate how to measure a PASTIS matrix on one contrast level, and use the derived tolerance limits on a better performing contrast level. Finally, we intend to explore closed-loop modal control with a HOWFS by using the optical modes from a measured PASTIS matrix, since these modes represent the direct sensitivity of the instrument contrast to segmented mirror misalignments.

\begin{acknowledgements}

IL thanks Lucie Leboulleux for discussions about the earlier work on the PASTIS tolerancing model. IL also thanks Ananya Sahoo for many discussions on the extension of this tolerancing model to thermo-mechanical modes and dynamical time scales of an AO system in space. The authors thank Michael Helmbrecht from IrisAO Inc. for the continued support and helpful discussions. The authors are especially thankful to the extended HiCAT team (over 50 people) who have worked over the past several years to develop this testbed. In particular, the authors thank Evelyn McChesney for the design and manufacturing of our deformable mirror mounts.
This work was co-authored by employees of the French National Aerospace Research Center ONERA (Office National d'\'{E}tudes et de Recherches A\'{e}rospatiales), and benefited from the support of the WOLF project ANR-18-CE31-0018 of the French National Research Agency (ANR). This work was supported by the Action Spécifique Haute Résolution Angulaire (ASHRA) of CNRS/INSU co-funded by CNES. This work was also supported in part by the National Aeronautics and Space Administration under Grant 80NSSC19K0120 issued through the Strategic Astrophysics Technology/Technology Demonstration for Exoplanet Missions Program (SAT-TDEM; PI: R. Soummer). E.H.P is supported by the NASA Hubble Fellowship grant \#HST-HF2-51467.001-A awarded by the Space Telescope Science Institute, which is operated by the Association of Universities for Research in Astronomy, Incorporated, under NASA contract NAS5-26555.

This research was developed in Python, an open source programming language. HiCAT makes use of the Numpy \citep{Oliphant2006numpy, vanderWalt2011NumPyArrayStructure}, Matplotlib \citep{Hunter2007matplotlib, matplotlib_v3.3.3}, Astropy \citep{AstropyCollaboration2013, AstropyCollaboration2018, Astropy2018zenodo}, SciPy \citep{scipy}, scikit-image \citep{scikit-image}, pandas \citep{Mckinney2010pandas, pandas_v1.1.4}, imageio \citep{imageio}, photutils \citep{photutils_v.1.0.1}, HCIPy \citep{Por2018HighContrastImaging}, Poppy \citep{Perrin2012poppy} and CatKit \citep{catkit0.36.1} packages. The software for the tolerancing analysis used the PASTIS \citep{Laginja2021pastis} Python package.
 
\end{acknowledgements}

\bibliographystyle{aa}
\bibliography{2021experimental}

\end{document}